\documentclass[11pt]{article}

\usepackage[preprint]{acl}

\usepackage{times}
\usepackage{latexsym}

\usepackage[T1]{fontenc}

\usepackage[utf8]{inputenc}

\usepackage{microtype}
\usepackage{inconsolata}

\usepackage{graphicx}
\usepackage{booktabs}
\usepackage{pifont}
\usepackage{amsmath}
\usepackage{amssymb}
\usepackage{tabularx}
\usepackage{longtable}
\usepackage{array}
\usepackage{xcolor}

\newif\ifrevmark
\revmarkfalse
\definecolor{revgreen}{RGB}{0,140,60}
\newcommand{\revise}[1]{\ifrevmark{\color{revgreen}#1}\else{#1}\fi}
\newcommand{\revisefloat}{\ifrevmark\color{revgreen}\fi}

\newcommand{\ok}{\ensuremath{\checkmark}}
\newcommand{\no}{\ensuremath{\times}}
\newcommand{\IntExt}[2]{\ensuremath{\mathrm{Int}#1/\mathrm{Ext}#2}}

\title{VeriEvolve: Evaluator-First Evolution for Tool-Using Agents}

\author{
  Yueshen Li\textsuperscript{1} \quad
  Kamer Ali Yuksel\textsuperscript{2} \quad
  Gokhan Tur\textsuperscript{1} \\
  \textbf{Dilek Hakkani-Tür\textsuperscript{1}} \quad
  \textbf{João Pedro Maia\textsuperscript{2}} \\
  \textsuperscript{1}University of Illinois Urbana-Champaign \quad
  \textsuperscript{2}aiXplain
}

\begin{document}
\maketitle

\begin{abstract}
A central obstacle to self-improving agents is the absence of a reliable fitness signal for realistic stateful tasks. We introduce \textsc{VeriEvolve}, an evaluator-first evolution framework that constructs this missing evaluation layer before optimizing the agent. 
\revise{From each task instruction}, VeriEvolve synthesizes instruction-grounded challenge states and an executable evaluator that checks them. 
During each round of workflow evolution,
the evaluator is frozen and serves as the fitness signal. Between rounds,
VeriEvolve revises the evaluator by converting the end states of strong and weak runs into a pair of counterfactual test cases.

Across all 30 MCPMark Filesystem tasks, VeriEvolve constructs 30 evaluators with 141 challenge states, achieving 141/141 agreement on the generated suites and 113/122 agreement on held-out states produced by executor agents. Evaluator-guided evolution increases externally verified success from 6/30 to 18/30 for GLM-5.2 and from 27/30 to 30/30 for Claude; refinement removes all three false positives that the weak executor exposes. Cross-environment studies on PostgreSQL, Playwright, Notion, and GitHub
further expose complementary limits from observability, state representation, and search/execution capacity. These results show that executable evaluators can be synthesized and used to guide search when task success is observable and an external verifier is available for calibration.
\end{abstract}

\section{Introduction}

A self-improving agent can optimize only what its evaluator can reliably
measure. In program synthesis, formal reasoning, and scientific optimization,
candidate quality is often supplied through unit tests, proof checkers, or
explicit numerical objectives
\citep{MathematicalDiscoveriesProgram,novikovAlphaEvolveCodingAgent,zhang2025DarwinGodelMachine,wang2025HuxleyGodelMachineHumanLevel}.
Realistic tool-use workflows rarely come with an equivalent fitness function.
Instead, agents must translate natural-language instructions into structured
changes across filesystems, databases, repositories, productivity platforms,
and browsers. Extending self-improvement to these settings therefore requires
not only better search but also the scalable construction of deterministic,
machine-checkable evaluators.

This evaluation bottleneck is especially pronounced in stateful,
create--read--update--delete (CRUD) workflows. Correctness is determined by the
resulting environment state rather than the final natural-language response,
and may depend jointly on artifact existence, content, hierarchy, consistency,
preservation, and the absence of unintended side effects. These constraints are
implied by the instruction but are generally not supplied as a complete
executable specification. Human-authored verification scripts can represent
them reliably, but they are costly to design and validate; MCPMark reports several
hours of expert effort for constructing each task and its verifier
\citep{wu2025MCPMarkBenchmarkStressTestinga}. LLM-based judges are easier to
scale, but semantic assessment alone cannot reliably establish that \revise{all required
properties of the final state hold}.

We study whether this missing evaluation layer can itself be constructed by
agents. Our central question is: \emph{Can an agent synthesize a
task-grounded executable evaluator that is \revise{informative enough} to guide
workflow improvement without access to an externally authored verifier during
search?} Our goal is not to reproduce an expert verifier perfectly in one step.
Instead, we ask whether an initially constructed evaluator can bootstrap useful
optimization, and whether the resulting execution states can subsequently
reveal missing constraints and support sharper evaluation artifacts.
\revise{We treat this as a feasibility question in a deliberately favorable
setting: task success is observable through the environment state, and an
independent benchmark verifier exists, which we use only outside the search
loop: to admit evaluators, identify false positives, and measure outcomes.}

We introduce \textsc{VeriEvolve}, an evaluator-first framework with four
model-backed roles: a Test-Suite Generator, Verifier Evolver, Workflow
Evolver, and Workflow Executor. In Stage~1, the framework derives positive,
negative, and adversarial challenge states from the task instruction and
observable initial environment. Deterministic category adapters materialize
these specifications, and the Verifier Evolver constructs a task-specific
executable evaluator that is frozen after passing internal gates. In Stage~2,
this evaluator supplies a stationary fitness signal for evolving a
tool-use workflow whose nodes retain persistent behavioral alternatives.
Stage~3 then compares high-confidence strong-executor states with
internally accepted weak-executor states, mines instruction-grounded
differences into paired counterfactual cases, and promotes a revised evaluator
only after cumulative regression. The revised evaluator is frozen before
workflow search resumes. \revise{The external MCPMark verifier is never shown
to a model-backed role and never ranks workflows or edits an evaluator. It is
used outside the loop for three purposes: admitting evaluators to Stage~2,
identifying which accepted states are false positives and therefore enter
Stage~3, and measuring results.}

Our primary controlled study covers all 30 MCPMark Filesystem tasks. Stage~1
constructs \revise{30/30 evaluators with 141 challenge states}, including
62 adversarial near-misses, with 141/141 agreement against the external
verifier on the generated suites. \revise{On 122 distinct states later produced
by the executor agents, which no evaluator saw during construction, agreement
is 113/122.} Under the weaker GLM-5.2 executor, workflow
evolution increases externally verified success from 6/30 at round~0 to
18/30; under Claude, it increases success from 27/30 to 30/30.
Three internally accepted GLM-5.2 states nevertheless fail external
verification, \revise{showing} that an evaluator can guide effective search while
remaining incomplete. In all three cases, strong--weak state differences yield
regression-verified evaluator revisions that eliminate the original false
positives; for \texttt{structure\_mirror}, \revise{the weak executor reaches an externally
valid state after a single revision.} An expanded PostgreSQL study covers all 21 standard tasks: 15
synthesized evaluators meet the Stage~2 \revise{admission threshold}, and workflow
evolution closes three weak-executor gaps that the strong executor solves at
round~0, while \revise{neither executor reaches internal convergence on the
remaining 12 eligible tasks within the configured limits}. Playwright, Notion,
and GitHub stress tests further expose complementary limits imposed by
observability and \revise{state representation}.

Our contributions are:
\begin{itemize}
    \item We \revise{identify the} scalable construction of deterministic,
    machine-checkable evaluators as a missing prerequisite for self-improving
    agents in realistic stateful tool-use workflows.

    \item We introduce \textsc{VeriEvolve}, which constructs task-grounded
    challenge states and executable evaluators before evolving persistent
    multi-alternative workflows under a frozen fitness signal\revise{, and
    revises an evaluator from differences between strong- and weak-executor
    final states}.

    \item \revise{On MCPMark, we show that the synthesized evaluators agree with
    the external verifier on their own challenge states and on held-out
    agent-produced states, that they guide externally measured workflow
    improvement for a strong and a weak executor, and that state-difference
    refinement removes all three false positives the external verifier
    exposed.}

    \item \revise{We will release the framework code and the resulting artifacts
    upon publication: challenge states and frozen evaluators for 30 Filesystem
    and 18 PostgreSQL tasks, the 258 agent-produced states with internal and external labels used
    for held-out evaluation, and the evolved workflows.}
\end{itemize}

\section{Related Work}

\paragraph{Self-improving agents and their evaluators.}
Language models are increasingly used as search operators over programs and
agents. FunSearch and AlphaEvolve evolve programs against automatically
computed objectives \citep{MathematicalDiscoveriesProgram,novikovAlphaEvolveCodingAgent};
the Darwin and Huxley G\"odel Machines evolve self-modifying coding agents
scored on coding benchmarks
\citep{zhang2025DarwinGodelMachine,wang2025HuxleyGodelMachineHumanLevel}; and
HyperAgents jointly represent task-solving and self-modification roles
\citep{zhang2026Hyperagents}. These systems start from a supplied benchmark or
utility. Recent work allows the evaluation criterion to change: the Red Queen
G\"odel Machine co-evolves agents with learned evaluators, freezing each
evaluator within an epoch and promoting challengers against held-out
ground-truth anchors \citep{iacob2026RedQueenGodel}; self-modifying Lean agents
co-evolve their benchmark while keeping Lean as a fixed trusted verifier
\citep{li2026SelfModifyingLeanProofa}; and the Verification Horizon argues that
verification must develop alongside increasingly capable generators
\citep{wang2026VerificationHorizonNo}. In each case, an improved evaluator is
still validated against an external benchmark, formal verifier, ground-truth
anchor, or human judgment. VeriEvolve shares this dependence: it builds
evaluators for stateful tool-use tasks without the benchmark's verifier but
uses that verifier afterward \revise{to admit evaluators, select false positives for
refinement, and measure their performance}.

\paragraph{Evaluation of stateful tool use.}
LLM judges scale feedback for open-ended text and dialogue
\citep{liu2023GEvalNLGEvaluation,zheng2023JudgingLLMasajudgeMTbench}, but a
semantic judgment cannot establish that every required state invariant holds.
Agent benchmarks therefore check the resulting artifacts or the environment state
programmatically
\citep{jimenez2023SWEbenchCanLanguage,yao2024TauBenchBenchmarkToolAgentUser,xie2024OSWorldBenchmarkingMultimodal,hattami2025WebArenaVerifiedReliable,wu2025MCPMarkBenchmarkStressTestinga},
and agent harnesses build on such executable feedback
\citep{ning2026CodeAgentHarnessa}. These checks are written by hand: MCPMark
reports several hours of expert effort per task and verifier
\citep{wu2025MCPMarkBenchmarkStressTestinga}. LLMs can generate tests and
checks from source code, formal specifications, or agent traces
\citep{wang2025TestEvalBenchmarkingLarge,cunha2026ValidatingFormalSpecifications,kohl2026AutomatedStructuralTestinga};
VeriEvolve instead synthesizes evaluators of final environment states from a
task instruction and its initial state, and revises them using states that
agents produce.

\section{Problem Setting}
\label{sec:problem}

\revise{A stateful tool-use task is a tuple \(\tau=(x,s_0,\mathcal{O})\): a
natural-language instruction \(x\), an initial environment state \(s_0\), and a
read-only observation interface \(\mathcal{O}\) that exposes the state available
for evaluation. Executing a workflow \(W\) with the task's tools yields a final
state \(s_W\). The correctness of \(s_W\) depends on constraints implied by the
instruction (existence, content, structure, consistency, and preservation of
artifacts), which are not given as an executable specification. We consider
only tasks whose success is observable through \(\mathcal{O}\).

\paragraph{Evaluator.}
An evaluator \(E\) is a deterministic program that maps \(\mathcal{O}(s)\) to a
verdict \(\hat{y}(s)\in\{0,1\}\), a partial score \(r(s)\in[0,1]\), and
check-level evidence. VeriEvolve builds \(E\) from \(x\), \(\mathcal{O}(s_0)\),
a set of synthesized challenge states with instruction-derived labels,
\(\mathcal{C}=\{(s_i,y_i)\}\), and deterministic execution feedback; it does not
use the benchmark's verifier. Workflow search maximizes \(r(s_W)\) while \(E\)
is held fixed, and a workflow is \emph{internally converged} when
\(\hat{y}(s_W)=1\).

\paragraph{Refinement.}
Evaluators are versioned \(E^{(0)},E^{(1)},\ldots\) together with their
challenge sets \(\mathcal{C}^{(k)}\). Refinement round \(k\) compares a reference
state \(F^{+}\) produced by a strong executor with a weak-executor state
\(F_k^{-}\) that \(E^{(k)}\) accepts. Their observable difference
\(\Delta^{(k)}=\mathrm{Diff}\bigl(\mathcal{O}(F^{+}),\mathcal{O}(F_k^{-})\bigr)\)
is turned into new labeled challenge states, giving \(\mathcal{C}^{(k+1)}\), and
into a revised evaluator \(E^{(k+1)}\); weak-executor search is then rerun.
\(F^{+}\) is treated as one high-confidence valid execution, not as a
ground-truth oracle, so a revision may add only instruction-grounded
distinctions and must preserve all earlier verdicts.}

\section{VeriEvolve Framework}
\label{sec:framework}

\begin{figure*}[t]
\centering
\IfFileExists{figures/framework_overview.pdf}{
    \includegraphics[width=0.97\textwidth]{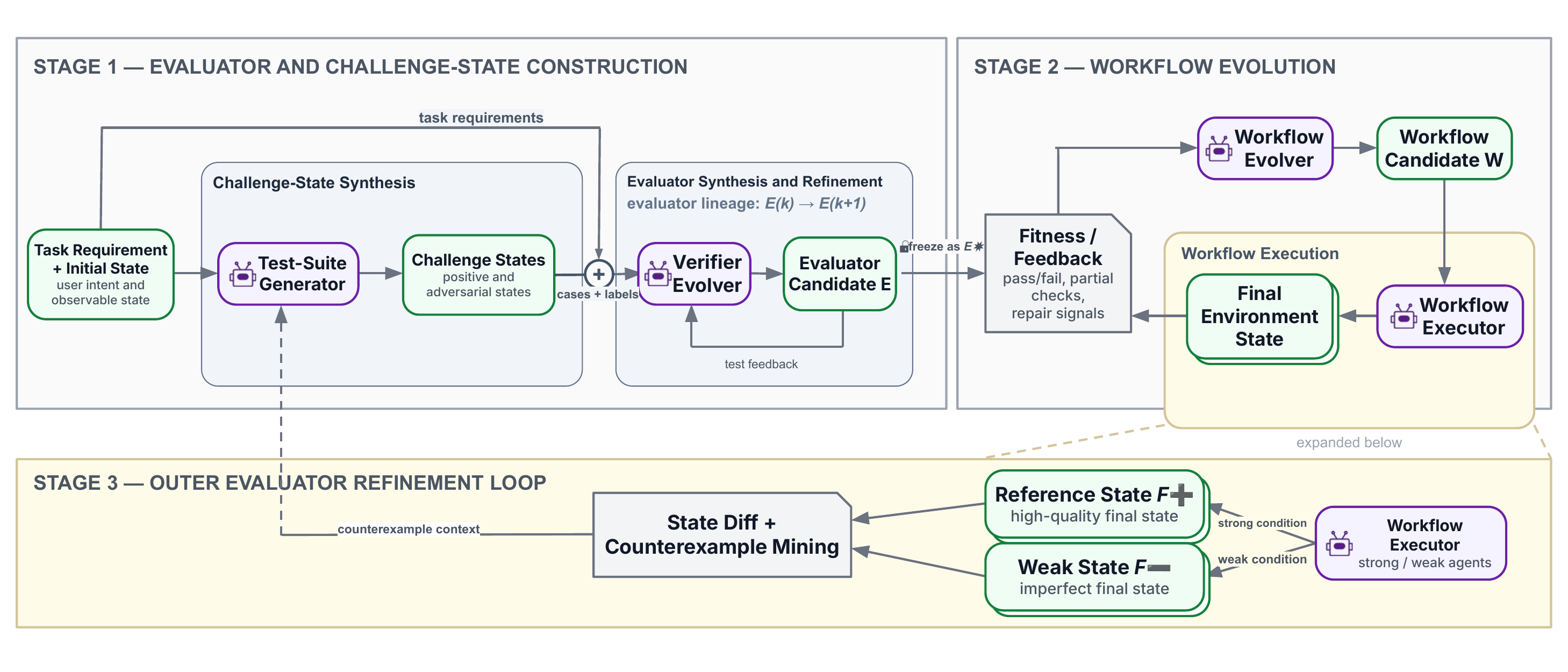}
}{
    \fbox{\parbox[c][1.55in][c]{0.94\textwidth}{\centering Placeholder for the VeriEvolve framework overview.}}
}
\caption{
Overview of VeriEvolve.
Stage~1 constructs challenge states and an executable evaluator; Stage~2 freezes
that evaluator while evolving workflows; and Stage~3 mines strong--weak
execution-state differences to extend the challenge set and revise the
evaluator.
One Stage~3 update followed by a Stage~2 rerun is a \revise{\emph{refinement round}}.
Purple nodes are model-backed roles, green nodes are persisted artifacts or
states, and gray nodes are deterministic operations.
}
\label{fig:framework}
\end{figure*}

Figure~\ref{fig:framework} operationalizes the objects defined in
Section~\ref{sec:problem}. Model-backed roles propose challenge states,
evaluator revisions, and workflow alternatives, while deterministic components
materialize states, execute and gate evaluators, compute state differences, and
version the resulting artifacts. The evaluator is frozen within each
workflow-search run and may be revised only between runs. \revise{Figure~\ref{fig:structure_mirror}
shows a task example following all three stages.}

\subsection{\texorpdfstring{\revise{Evaluator Construction}}{Evaluator Construction}}
\label{sec:stage1}

The Test-Suite Generator maps the task instruction and observable initial state
to declarative positive, clean-negative, and adversarial challenge cases. A
category-specific deterministic adapter materializes each specification in an
isolated copy of the task environment, producing executable states with
instruction-derived labels. 

The Verifier Evolver constructs a sandboxed executable evaluator over the
read-only observation interface \(\mathcal{O}\). A deterministic harness runs
each evaluator candidate on the materialized challenge states and returns
case-level failures as refinement feedback. Once the candidate satisfies the
construction gate, it is frozen as \(E^{(0)}\). The initial output of Stage~1
is therefore the \revise{challenge set and evaluator}
\[
\left(
\mathcal{C}^{(0)},
E^{(0)}
\right).
\]

\subsection{Workflow Evolution under Frozen Evaluation}
\label{sec:stage2}

A workflow contains five functional nodes for task understanding, planning,
ReAct execution, artifact inspection, and post-processing. Each node maintains
a pool of prompt or behavior alternatives\revise{, following the persistent
multi-alternative representation of EvoLattice \citep{yuksel2025EvoLatticePersistentInternalPopulation}}, and a candidate path selects one
alternative from every active node. The Workflow Executor runs the selected
path in an isolated task environment and returns the final state and execution
trace.

The frozen evaluator \(E^{(k)}\) scores the observable final state and provides
partial fitness and check-level evidence. Artifact inspection separately
summarizes execution behavior that may explain the failed checks. The Workflow
Evolver uses these signals to append new node alternatives for subsequent
rounds. Because earlier alternatives remain available, promising partial
behaviors can be retained and recombined rather than overwritten by each
mutation. Throughout a Stage~2 run, however, \(E^{(k)}\) is immutable; the
search changes how the task is understood, planned, executed, inspected, or
post-processed, but not the criterion used to compare candidates. \revise{The
multi-alternative workflow representation is shown in Appendix~\ref{app:workflow_repr}
(Figure~\ref{fig:workflow}).}

\subsection{Execution-State Refinement}
\label{sec:state_refinement}

\paragraph{State-difference mining.}
At \revise{refinement round} \(k\), a deterministic adapter first removes known
\revise{environment-specific} artifacts symmetrically from \(F^{+}\) and \(F_k^{-}\).
The State-Diff module then organizes the remaining observable differences into
factual clusters, such as missing or extra artifacts, content mismatches, and
structural deviations. Only clusters that can be grounded in the task
instruction are eligible for refinement.

\paragraph{Counterfactual challenge states.}
For each selected cluster, the Test-Suite Generator constructs paired
counterfactual states using \(F^{+}\) and the current \(F_k^{-}\) as immutable
anchors. The negative state reproduces the selected difference, while the
positive control removes it and holds the unrelated state fixed. A deterministic
label-consistency gate verifies the expected verdicts, and a coverage gate
checks that the pair actually introduces and removes the targeted difference.
Cases that fail either gate are rejected before evaluator refinement.

\paragraph{Evaluator promotion and workflow rerun.}
The Verifier Evolver receives the current evaluator, the cumulative challenge
set, and factual observation-level contrasts between the paired states. It
proposes a localized revision targeted at the uncovered constraint. A candidate
is promoted to \(E^{(k+1)}\) only if cumulative regression verifies that it
accepts \(F^{+}\), rejects all prior weak false-positive states, preserves all
previous challenge-state verdicts, and correctly classifies the new cases.
The promoted evaluator is then frozen, and weak-executor workflow evolution is
rerun under the sharper criterion.

A further \revise{refinement round} is useful only when the new weak state exposes a novel,
instruction-grounded difference that the current evaluator does not already
cover. VeriEvolve terminates when the weak state matches the reference or no
actionable difference remains; when the evaluator stabilizes on a repeated,
already-covered difference; or when workflow search, evaluator refinement, or
the \revise{refinement-round} budget is exhausted. Repeated production of an error already
detected by the evaluator is therefore attributed to workflow-search or
executor insufficiency rather than treated as another evaluator gap.

Each accepted \revise{refinement round} appends paired challenge states, a strong--weak state
contrast, and a new evaluator version, yielding an ordered and reusable
\revise{evaluator lineage}. The external benchmark verifier \revise{is not part of
this loop: its source, verdicts, and diagnostics are never provided to any
model-backed role or used for workflow ranking, evaluator repair, or
termination. In our experiments, however, external verdicts decide which
evaluators enter Stage~2 and which internally accepted states are false
positives that trigger refinement, and the reference state \(F^{+}\) is one
that the external verifier also accepts. A deployment without an external
verifier would need another trigger and reference. We evaluate one candidate
trigger, executor disagreement, only retrospectively
(Appendix~\ref{app:trigger}); a verifier-free reference is not evaluated.}
\section{Experimental Setup}
\label{sec:experimental_setup}

\subsection{Evaluation Questions and Scope}

We evaluate on a subset of the MCPMark standard task suite spanning five task
environments; the complete task and exclusion breakdown is reported in
Appendix~\ref{app:task_scope}. We treat Filesystem as our primary controlled
study and the remaining environments as cross-environment stress tests.

\revise{We ask three questions, one per stage. \textbf{RQ1} (evaluator
construction): can VeriEvolve synthesize challenge states and executable
evaluators that distinguish correct task states from minimally incorrect or
adversarial ones? \textbf{RQ2} (workflow evolution): do the synthesized
evaluators give feedback informative enough to guide workflow search?
\textbf{RQ3} (execution-state refinement): can differences between strong- and
weak-executor states reveal evaluator constraints not yet covered?}

\subsection{Models and Experimental Budgets}

\revise{All model-backed roles use Claude Sonnet, except that initial Stage~1
construction ran in interactive Claude Code sessions and the weak Workflow
Executor is GLM-5.2; Table~\ref{tab:appendix_model_roles} in
Appendix~\ref{app:experimental_details} gives identifiers and invocation
modes.}

Stage~1 evaluator construction uses only internal gates: positive acceptance of
at least \(0.90\) and adversarial rejection of at least \(0.80\). After an
evaluator is frozen, we measure its agreement with the external MCPMark
verifier. For the Stage~2 \revise{study}, we include frozen evaluators
with at least \(0.85\) post-hoc external agreement. This external measurement
\revise{determines study inclusion and is used for evaluation; Stage~3 is applied
to the internally accepted states that the external verifier rejects. External}
labels, diagnostics, and verifier logic are never routed back into test
generation, evaluator repair, workflow ranking, or workflow evolution.

Unless otherwise noted, \revise{Stage~2} runs use five paths \revise{per round (three for
the Filesystem Claude campaign)}, at most 15 workflow rounds \revise{(three paths and
at most five rounds for the Playwright and Notion pairs)}, up to three evaluator-repair attempts, and at most three refinement
\revise{rounds}. \revise{In the Filesystem GLM-5.2 and the PostgreSQL campaigns, each Stage~2
round ran as a subprocess with a one-hour timeout; a round that exceeded it
ended the run, which is counted as unsolved.}
Full per-category gates, budgets, and timeout settings are listed
in Appendix~\ref{app:experimental_details}.

We denote acceptance by the internal evaluator and external MCPMark verifier
as \(\mathrm{Int}\checkmark\) and \(\mathrm{Ext}\checkmark\), respectively.
Thus, \IntExt{\ok}{\no} denotes an evaluator false
positive and \IntExt{\no}{\ok} an evaluator false
negative. External verdicts are \revise{never} used as optimization feedback. Formal exclusions and
infrastructure-invalid runs are reported separately and are not counted as
semantic task failures.
\section{Results}
\label{sec:results}

\subsection{RQ1: Constructing \texorpdfstring{\revise{Evaluators}}{Evaluators}}
\label{sec:results_rq1}

\begin{table}[t]
    \centering
    \scriptsize
    \setlength{\tabcolsep}{2.6pt}
    \begin{tabular}{lcccc}
        \toprule
        Environment
        & Constructed
        & Cases (P/N)
        & Ext.\ agreement
        & Stage-2 study \\
        \midrule
        Filesystem
        & 30/30
        & 141 (34/107)
        & 141/141
        & 30/30 \\
        PostgreSQL
        & 18/21
        & 120 (22/98)
        & 115/120
        & 15/21 \\
        \bottomrule
    \end{tabular}
    \caption{
    Primary RQ1 evidence. Cases are generated challenge states; P/N reports
    their post-hoc external-verifier labels. External agreement compares the
    frozen internal evaluator with those labels. The external verifier is
    consulted only after evaluator construction and is not exposed to any
    model-backed role or used as evaluator-repair feedback. For the Stage~2
    \revise{study}, we include tasks whose frozen evaluator reaches at least
    \(0.85\) post-hoc external agreement.
    }
    \label{tab:rq1_packages}
\end{table}

Table~\ref{tab:rq1_packages} shows that VeriEvolve constructs \revise{evaluators} across two different state interfaces. For the full
Filesystem study, Stage~1 constructs \revise{30/30 evaluators with} 141 materialized
challenge states, including 62 adversarial near-misses. After the evaluators
are frozen, post-hoc calibration finds agreement with the external verifier on
all 141 states.

For PostgreSQL, VeriEvolve successfully constructs evaluators for \revise{18/21}
standard tasks, covering 120 calibration states (22 positive and 98 negative).
Post-hoc calibration of the frozen evaluators yields 115/120 agreement with the
external verifier. Fifteen evaluators reach at least \(0.85\) per-task
agreement and are included in the Stage~2 \revise{study}; three frozen
evaluators remain below this study-inclusion threshold, while the remaining
three tasks are \revise{blocked by the observation interface} because their success conditions are not
adequately represented by the available observation interface. Importantly,
external verdicts are obtained only after evaluator construction: they
determine the scope of the downstream \revise{study} but are never
routed back into test generation, evaluator repair, workflow ranking, or
workflow evolution.

\begin{table}[t]
    \revisefloat
    \centering
    \scriptsize
    \setlength{\tabcolsep}{2.2pt}
    \begin{tabular}{@{}lrrrrrrrrr@{}}
\toprule
 & & \multicolumn{4}{c}{Int/Ext} & & & & \\
\cmidrule(lr){3-6}
Executor & States & \ok/\ok & \ok/\no & \no/\ok & \no/\no & Agr. & Prec. & Rec. & Spec. \\
\midrule
GLM-5.2 & 85 & 18 & 3 & 1 & 63 & 95.3 & 85.7 & 94.7 & 95.5 \\
Claude & 44 & 30 & 0 & 5 & 9 & 88.6 & 100.0 & 85.7 & 100.0 \\
\midrule
All (distinct) & 122 & 41 & 3 & 6 & 72 & 92.6 & 93.2 & 87.2 & 96.0 \\
\bottomrule
\end{tabular}

    \caption{Held-out evaluation on Filesystem: frozen Stage~1 evaluators
    versus the external verifier on distinct agent-produced states from both
    Stage~2 campaigns (Int/Ext columns count states; Agr., Prec., Rec., and
    Spec.\ are agreement, precision, recall, and specificity in \%, with
    ``pass'' as the positive class). Seven states produced by both executors are
    counted once in the last row. Details in Appendix~\ref{app:heldout}.}
    \label{tab:heldout}
\end{table}

\revise{\paragraph{Held-out agent-produced states.}
Agreement on a finite generated suite does not imply complete coverage. To
test the frozen evaluators beyond their own challenge suites, we re-execute
each Stage~1 evaluator on the final states that the two Stage~2 executors
produced on Filesystem and label every state post hoc with the external
verifier. These states were produced after the evaluators were frozen, by other
models working on the real task. The campaigns retained the best state of each
round, giving 258 states, 122 of them distinct (Table~\ref{tab:heldout}). The
evaluators agree with the external verifier on 113/122 distinct states
(92.6\%; 95\% CI 86.6--96.1). Errors occur in both directions: the three false
positives are the states that later trigger Stage~3, and six valid states, on
four tasks, are rejected internally. Restricted to the 60 final states, agreement
is 57/60, the outcomes in Table~\ref{tab:rq2_workflow}. Because each retained
state is the one the evaluator ranked highest in its round, this pool
over-represents internally high-scoring states and is not an independent
sample. The smaller live-state studies show the same scope boundary
(Appendix~\ref{app:category_results}).}

\paragraph{Answer to RQ1.}
\revise{Where correctness is observable, VeriEvolve built evaluators that agree
with the external verifier on their own suites (141/141 and 115/120) and on
113/122 held-out Filesystem states. This is external calibration, not evidence
of reliability on its own.}

\subsection{RQ2: Guiding Workflow Evolution}
\label{sec:results_rq2}

\begin{table}[t]
    \centering
    \scriptsize
    \setlength{\tabcolsep}{2.4pt}
    \begin{tabular}{@{}llrrrr@{}}
        \toprule
        Executor
        & Workflow
        & Int.\ pass
        & Ext.\ pass
        & FP
        & FN \\
        \midrule
        Claude
        & Seed
        & 24/30
        & 27/30
        & 0
        & 3 \\
        Claude
        & Evolved
        & 30/30
        & 30/30
        & 0
        & 0 \\
        GLM-5.2
        & Seed
        & 8/30
        & 6/30
        & 2
        & 0 \\
        GLM-5.2
        & Evolved
        & 21/30
        & 18/30
        & 3
        & 0 \\
        \bottomrule
    \end{tabular}
    \caption{
    Filesystem seed and evolved workflow performance.
    Seed denotes the internally selected best candidate at round~0, before
    workflow evolution, under each campaign's recorded path budget.
    FP and FN denote
    \IntExt{\ok}{\no} and
    \IntExt{\no}{\ok}, respectively.
    }
    \label{tab:rq2_workflow}
\end{table}

On Filesystem, workflow evolution triples the weak executor's externally
verified success from 6/30 to 18/30 and raises internal convergence
from 8/30 to 21/30. Thus, 13 tasks reach internal convergence only
after workflow evolution. Of the 21 final internally accepted states, 18 pass
external verification, giving \(85.7\%\) internal precision. The three
remaining false positives become the execution-state refinement tasks in RQ3;
no GLM-5.2 state is externally accepted while internally rejected.

The strong Filesystem executor improves from 27/30 to 30/30 external
success, while internal convergence rises from 24/30 to 30/30. Its
three round-0 disagreements are evaluator false negatives, and no
strong-executor state is a false positive. The strong condition also
\revise{records} approximately \(58\times\) fewer executor tokens than GLM-5.2 \revise{over the
full campaign},\footnote{\revise{Claude executor counts are input plus output tokens as
reported by the Claude CLI and exclude prompt-cache reads; the ratio is
therefore indicative rather than exact.}} highlighting the remaining effect of
executor capability under the same task-specific evaluators.
\revise{Because both comparisons are paired over the same 30 tasks, we apply an
exact McNemar test (Table~\ref{tab:fs_significance}): GLM-5.2's external gain
(12 tasks gained, none lost) is significant (\(p<0.001\)), whereas Claude's
three gained tasks are not (\(p=0.25\)).}

\revise{On the 15 admitted PostgreSQL tasks, both executors solve the same three
tasks, and every internally accepted final state is also externally accepted.
The strong executor solves them at round~0; the weak executor needs 11, 2, and
7 rounds of evolution and 5--32\(\times\) the strong executor's tokens. Neither
executor converges on the other 12 tasks within the configured limits
(Appendix~\ref{app:category_results}), which also reports the smaller
Playwright and Notion studies.}

\paragraph{Answer to RQ2.}
\revise{The evaluators give usable search feedback: evolution triples
weak-executor success on Filesystem and closes three PostgreSQL gaps, but
12/15 PostgreSQL tasks stay unsolved, so a usable evaluator does not remove
executor and search limits.}

\subsection{RQ3: Refining Evaluators from Execution States}
\label{sec:results_rq3}

\begin{figure*}[t]
    \centering
    \includegraphics[width=0.95\textwidth]{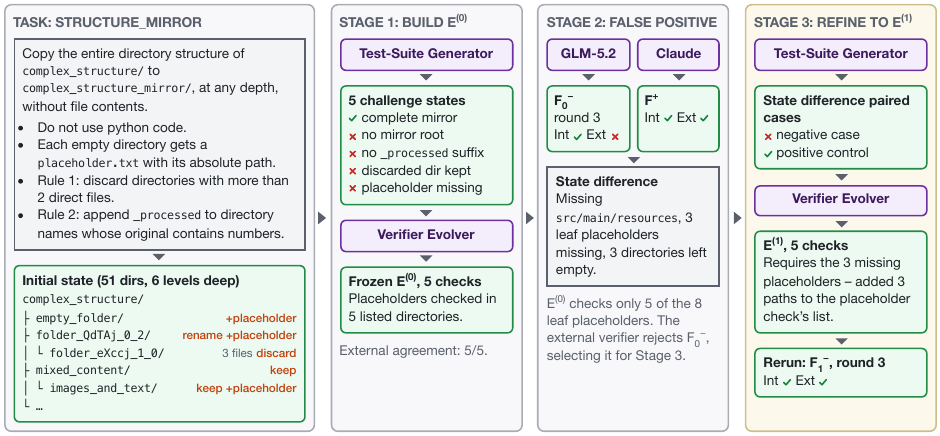}
    \caption{\revise{Worked example on the MCPMark Filesystem task
    \texttt{structure\_mirror}. Orange labels give what the mirror must do
    with each directory. Stage~1 builds \(E^{(0)}\) from five challenge states;
    \(E^{(0)}\) checks placeholders in only 5 of the 8 empty leaf directories.
    In Stage~2, GLM-5.2 reaches a state \(F_0^{-}\) that \(E^{(0)}\) accepts
    but the external verifier rejects, which selects the task for Stage~3.
    Stage~3 turns the difference from the Claude state \(F^{+}\) into paired
    cases and a revised evaluator \(E^{(1)}\), under which the weak
    executor's next state \(F_1^{-}\) passes both checks. Int and Ext denote
    the internal evaluator and the external verifier.}}
    \label{fig:structure_mirror}
\end{figure*}

The final weak-executor Filesystem campaign produces exactly three
\IntExt{\ok}{\no} states:
\texttt{uppercase}, \texttt{debugging}, and \texttt{structure\_mirror}.
We apply one task-generic refinement strategy to all three, starting from their
original evaluator, weak state, and workflow. \revise{These three cases are
selected by the external verifier, and each reference state \(F^{+}\) is a
Claude state that it also accepts; the refinement itself receives only the
instruction, the two states, and the current evaluator and challenge set.}
\revise{The repaired constraints are exact text and terminal-newline fidelity
(\texttt{uppercase}), presence of the required fixed code pattern
(\texttt{debugging}), and the missing nested directories and placeholders
(\texttt{structure\_mirror}); Table~\ref{tab:rq3_refinement} in
Appendix~\ref{app:full_results} lists the revisions.}

Every strong--weak difference yields instruction-grounded paired challenge
states and a regression-free evaluator revision. The promoted evaluators
preserve the strong reference and all prior challenge-state verdicts while
rejecting every previously accepted weak state. Consequently, all three
original evaluator gaps are repaired and no
\IntExt{\ok}{\no} state survives the \revise{primary refinement
runs}. \revise{On \texttt{uppercase} and \texttt{debugging}, the trajectory
stops once the weak executor repeats an already-covered failure rather than
exposing another evaluator constraint.}

\revise{Figure~\ref{fig:structure_mirror} follows \texttt{structure\_mirror} end
to end. \(E^{(0)}\) lists only 5 of the 8 empty leaf directories that need a
placeholder, so it accepts a weak state that omits the other 3. The difference
from \(F^{+}\) forms three clusters (missing directories, missing placeholders,
and directories left empty); the six paired cases lead to a three-line
revision that adds the missing paths, and \(E^{(1)}\) passes cumulative
regression. Rerunning weak-executor search under \(E^{(1)}\), GLM-5.2
converges at round~3 on a state that the external verifier also accepts. The
run then entered a second refinement round, but its only observed difference
was a duplicate directory created by file synchronization after the state had
been archived; we exclude that round (Appendix~\ref{app:structure_mirror}).}

\paragraph{Answer to RQ3.}
\revise{Once the external verifier flags a false positive, strong--weak state
differences repair the missing constraint without further external feedback
(3/3 on Filesystem). One revision let the weak executor solve
\texttt{structure\_mirror}, but not \texttt{uppercase} or \texttt{debugging}.
Refinement needs an agent-produced false positive, which the PostgreSQL
campaign never yields.}

\subsection{\texorpdfstring{\revise{Cost and Repair Effort}}{Cost and Repair Effort}}
\label{sec:results_cost}

\begin{table}[t]
    \revisefloat
    \centering
    \scriptsize
    \setlength{\tabcolsep}{2.4pt}
    \begin{tabular}{@{}lrrr>{\raggedright\arraybackslash}p{0.33\linewidth}@{}}
\toprule
Run & Exec.\ tok. & Evolver & Wall & Normalized cost \\
\midrule
\multicolumn{5}{@{}l}{\emph{Filesystem}} \\
Stage~2, GLM-5.2 & 33.18M & \$31.24 & 43.5\,h & 2.76M, \$2.60 per gained task (12) \\
Stage~2, Claude & 0.58M$^{a}$ & \$1.74 & 5.3\,h & 0.19M, \$0.58 per gained task (3) \\
Stage~3 & 8.10M & \$3.49 & 4.6\,h$^{b}$ & 2.70M + 21.4K TSG/VE, \$1.16 per repaired evaluator (3) \\
\multicolumn{5}{@{}l}{\emph{PostgreSQL (15 tasks)}} \\
Stage~2, GLM-5.2 & 27.82M & \$22.74 & 37.4\,h & 9.27M per solved task (3) \\
Stage~2, Claude & 19.08M & \$19.14 & 33.9\,h & 6.36M per solved task (3) \\
\bottomrule
\end{tabular}

    \caption{Recorded cost of the main campaigns. Evolver is the Workflow
    Evolver's dollar cost; wall time is summed over tasks or runs. Normalized
    cost divides the whole campaign by the tasks that pass externally only
    after evolution (Filesystem), by repaired evaluators (Stage~3), or by
    externally verified successes (PostgreSQL). \(^{a}\)Excludes prompt-cache
    reads. \(^{b}\)2.5\,h elapsed with runs in parallel. Stage~1 cost was not
    recorded.}
    \label{tab:cost}
\end{table}

\revise{Table~\ref{tab:cost} summarizes recorded cost (full accounting in
Appendix~\ref{app:compute}). Executor tokens dominate. The weak executor needs
2.76M executor tokens and \$2.60 of evolver cost for each Filesystem task it
gains through evolution, about 15 times the strong executor's executor tokens
and 4.5 times its evolver cost per gained task. The evaluator repair itself is cheap: across the three Stage~3 tasks, the
Test-Suite Generator was called four times and passed the label and coverage
gates on the first attempt each time, and the Verifier Evolver made six
attempts, of which two (both on \texttt{debugging}) were rejected by cumulative
regression. These calls use 64K tokens, under 1\% of Stage~3 tokens; the rest
re-runs weak-executor search under each revised evaluator (49 rounds in
total, including 15 rounds and 3.36M executor tokens spent in the excluded
\texttt{structure\_mirror} round).}

\section{Discussion}
\label{sec:discussion}

\revise{Once a calibrated task-specific evaluator is available, the bottleneck can
shift from evaluation to search and execution: neither executor converges on 12
of the 15 admitted PostgreSQL tasks, and on the three solved tasks only the weak
executor needs evolution. A usable fitness signal is therefore necessary but not
sufficient. Refinement is also trajectory-dependent: it starts only when search
reaches an evaluator-accepted state that exposes a gap, and in our experiments
that trigger came from the external verifier. A simple verifier-free
alternative, flagging any accepted weak-executor state that differs from the
strong executor's accepted state, would in retrospect have caught all three
false positives but also flagged 10 of 18 correct states
(Appendix~\ref{app:trigger}); making such a trigger precise remains open. Because compute is
dominated by candidate execution, future work should devote more capacity to
search and execution, while the accumulated challenge states and evaluator
revisions may support smaller models specialized for test generation and
evaluator repair.}

\section{Conclusion}

We introduced \textsc{VeriEvolve}, an evaluator-first framework that constructs
challenge states and executable evaluators before optimizing tool-use
workflows. \revise{On MCPMark tasks whose success is observable through the
environment state, the synthesized evaluators agree with the external verifier
on their own challenge states and on most held-out agent-produced states, guide
externally measured workflow improvement, and can be repaired from strong--weak
state differences once a false positive is identified. Throughout, the external
verifier remains the reference that admits evaluators and exposes their errors;
removing that dependence is the main step toward agents that construct the
fitness signals they require.}

\section*{Limitations}
\label{sec:limitations}

\revise{\paragraph{Dependence on an external verifier.}
VeriEvolve never exposes the benchmark verifier to a model or uses it inside
search, but our experiments are not verifier-free. External agreement decides
which evaluators enter Stage~2, external verdicts identify the false positives
that Stage~3 refines, and each reference state \(F^{+}\) is externally accepted.
Our results therefore show feasibility under external calibration, not that
synthesized evaluators are reliable, or that refinement can run autonomously,
without an independent verifier.}

\revise{\paragraph{Evidence for evaluator quality.}}
Calibration on a finite challenge suite does not guarantee complete evaluator
coverage. \revise{Test-suite generation and evaluator construction both used
Claude, so the 141 Stage~1 challenge states share their author with the
evaluators they test. The held-out agent-produced states come from other
executors, but only the best state of each round was retained, so they are
selected by the evaluator's own ranking and are not independent samples.}
Initial interactive Stage~1 construction additionally relied on a procedural
verifier-information boundary, whereas the stronger packet-level technical
isolation applies to Stage~3.

\revise{\paragraph{Coverage and search.}}
Our empirical coverage remains uneven across environments. Filesystem is the
only category with complete Stage~1 and Stage~2 coverage together with
execution-state refinement. PostgreSQL provides broader evidence: all 21
standard tasks receive a Stage~1 disposition, 15 evaluators qualify for
Stage~2, and all 15 matched strong/weak conditions are run to their configured
limits. However, only three of these tasks are solved by either executor, and
no PostgreSQL trajectory produces an \IntExt{\ok}{\no} state that would trigger
Stage~3. Playwright, Notion, and GitHub remain smaller or incomplete
cross-environment studies. VeriEvolve can evaluate only properties represented
through the observation interface, and several cross-environment tasks are
excluded or caveated because their success conditions are missing, weakly
specified, or distorted by the available state representation.

\revise{Each Filesystem condition was run once, and we compare evolved workflows
with their own round-0 seed rather than with a budget-matched search baseline,
so part of the measured gain may reflect repeated sampling rather than
evaluator guidance.} A usable fitness signal does not guarantee task completion: on most
PostgreSQL Stage~2 tasks neither executor converges\revise{, and several weak-executor
runs end at the one-hour round timeout rather than after all 15 rounds}.
Successful evaluator repair likewise does not guarantee that a weaker executor
can satisfy the refined criterion. Refinement requires a high-confidence
positive state \(F^{+}\) from a sufficiently capable executor, and \(F^{+}\) is
one valid execution, not a complete characterization of all acceptable
solutions. \revise{The strong executor provides such a state on 35 of
49 tasks across environments but on only 3 of 15 PostgreSQL tasks
(Appendix~\ref{app:strong_executor}); one of its internally accepted states
is rejected externally, and the effect of an erroneous \(F^{+}\) on the
revised evaluator is untested.} Repeated workflow execution and refinement remain computationally
expensive, limiting both the present experimental scale and the number of
opportunities for execution-state refinement.

\bibliography{custom}

\appendix
\appendix

\section{Task Scope and Evaluation Regimes}
\label{app:task_scope}

\subsection{Audited Experimental Scope}

Table~\ref{tab:appendix_scope} reports the complete experimental scope.
Formal exclusions are not counted as task failures, infrastructure-invalid
runs are excluded from semantic results, and historical or pilot runs are not
pooled with standardized campaigns.

\begin{table*}[t]
    \centering
    \scriptsize
    \setlength{\tabcolsep}{4pt}
    \begin{tabularx}{\textwidth}{lrrrrrX}
        \toprule
        Environment
        & Universe
        & RQ1 attempted
        & S2 eligible
        & Valid pairs
        & RQ3 cases
        & Disposition \\
        \midrule

        Filesystem
        & 30
        & 30
        & 30
        & 30-task campaigns
        & 3
        & Primary standardized study; no exclusions. \\

        PostgreSQL
        & 21
        & 21
        & 15
        & 15
        & 0
        & Full Stage~1 disposition and Stage~2 sweep over all eligible tasks;
          three additional evaluators fall below the \revise{admission threshold} and
          three tasks are \revise{blocked by the observation interface}. \\

        Playwright
        & 4
        & 4
        & 2
        & 2
        & 0
        & Two formal exclusions: missing independent ground truth and
          unobservable success state. \\

        Notion
        & 28
        & 23
        & 2
        & 1 historical
        & 0
        & One historical matched pair; broader standardized execution remains
          limited by the live-state \revise{representation}. \\

        GitHub
        & 23
        & 23
        & 1
        & 0 standardized
        & 0
        & Incomplete feasibility campaign; one externally calibrated evaluator
          and one corrected strong-executor \(F^{+}\), with the corrected
          weak-executor pair incomplete. \\
        \bottomrule
    \end{tabularx}
    \caption{
    Audited task scope. ``S2 eligible'' counts evaluators or tasks admitted to
    the corresponding Stage~2 study. Historical, diagnostic, and incomplete
    pilot results are reported separately from standardized matched pairs.
    }
    \label{tab:appendix_scope}
\end{table*}

\subsection{Explicitly Checkable and Implicit Stateful Requirements}

VeriEvolve targets tasks in which correctness is determined by a structured
environment state but a complete executable evaluation specification is not
provided to the agent. The difficulty of constructing that specification
varies substantially across tasks, as illustrated in
Table~\ref{tab:evaluation_regimes}.

\begin{table*}[t]
    \centering
    \small
    \setlength{\tabcolsep}{5pt}
    \begin{tabularx}{\textwidth}{llX}
        \toprule
        Task
        & Evaluation regime
        & Representative requirements \\
        \midrule
        \texttt{uppercase}
        & Nearly direct specification
        & Create explicitly named uppercase file copies and a fixed-format
          word-count artifact. Major instruction clauses map nearly
          one-to-one to deterministic evaluator checks. \\

        \texttt{structure\_mirror}
        & Implicit state specification
        & Reproduce a transformed hierarchy while satisfying interacting
          existence, naming, placeholder-content, discard, preservation, and
          absence constraints. The prohibition on invalid extra structure is
          exposed only through strong--weak execution-state comparison
          \revise{in the development and controlled runs (Appendix~\ref{app:structure_mirror})}. \\
        \bottomrule
    \end{tabularx}
    \caption{
    Examples of differing evaluation regimes within stateful tool-use tasks.
    Some instructions admit a nearly direct translation into executable
    checks, whereas others distribute correctness across interacting
    properties of the final environment state.
    }
    \label{tab:evaluation_regimes}
\end{table*}

\section{Implementation and Experimental Configuration}
\label{app:experimental_details}

\subsection{Model Roles and Invocation Modes}

Table~\ref{tab:appendix_model_roles} \revise{lists the model roles summarized in
Section~\ref{sec:experimental_setup}}. Initial Stage~1 construction used one
interactive Claude Code session in which the Test-Suite Generator and Verifier
Evolver acted sequentially. Stage~3 instead used a retained, headless session
per \revise{refinement round}: the Verifier Evolver resumed the accepted Test-Suite Generator
session under a packet-restricted interface. The Workflow Evolver was
stateless across rounds; the orchestration code explicitly supplied its
alternative pool, score history, failed checks, and execution diagnosis on
every call.

\revise{All Claude roles except initial Stage~1 were invoked through the Claude CLI
model alias \texttt{sonnet}; the recorded Stage~3 sessions resolve this alias
to \texttt{claude-sonnet-5}. The underlying model of the interactive Stage~1
sessions was not logged, so we report it only as Claude. The weak executor is
GLM-5.2 (\texttt{glm-5.2}) served through the Z.ai full-precision API endpoint.}

\begin{table*}[t]
    \centering
    \scriptsize
    \setlength{\tabcolsep}{3.5pt}
    \begin{tabularx}{\textwidth}{lllXX}
        \toprule
        Role
        & Base model
        & Invocation
        & Context and visible inputs
        & Output \\
        \midrule
        Test-Suite Generator, Stage~1
        & Claude
        & Interactive Claude Code CLI
        & Persistent task session; task instruction, observable initial state,
          and self-generated artifacts
        & Declarative labeled challenge-state specification \\

        Verifier Evolver, Stage~1
        & Claude
        & Same interactive session
        & Shares context with the Test-Suite Generator; receives challenge
          states and deterministic internal scoring feedback
        & Sandboxed \texttt{verify(context)} evaluator \\

        Test-Suite Generator, Stage~3
        & Claude Sonnet
        & Headless retained CLI session
        & Packet-restricted task instruction, current evaluator, sanitized
          cases, state-difference facts, and output contract
        & Anchor-based paired challenge cases \\

        Verifier Evolver, Stage~3
        & Claude Sonnet
        & Resumes Stage~3 TSG session
        & Current evaluator, cumulative cases, factual paired contrasts, and
          deterministic repair feedback
        & Localized evaluator edits or replacement code \\

        Workflow Evolver
        & Claude Sonnet
        & \texttt{claude -p}; one call per round
        & Fresh call with alternative statistics, score trajectory, failed
          checks, and artifact-inspection diagnosis
        & At most one new alternative per selected node \\

        Strong Workflow Executor
        & Claude Sonnet
        & Headless CLI subprocess per path
        & Candidate workflow prompts and isolated task environment
        & Final environment state and execution trace \\

        Weak Workflow Executor
        & GLM-5.2
        & API call per workflow node
        & Same workflow prompts and category-specific MCP tool surface
        & Final environment state and execution trace \\
        \bottomrule
    \end{tabularx}
    \caption{
    Model roles and invocation modes. Stage~1 TSG and VE share one interactive
    session; Stage~3 TSG and VE share one retained packet-restricted session per
    \revise{refinement round}. The Workflow Evolver and both executors are invoked separately
    for each round or candidate.
    }
    \label{tab:appendix_model_roles}
\end{table*}

\subsection{Persistent Multi-Alternative Workflow}
\label{app:workflow_repr}

\begin{figure*}[t]
    \centering
    \includegraphics[width=0.96\textwidth]{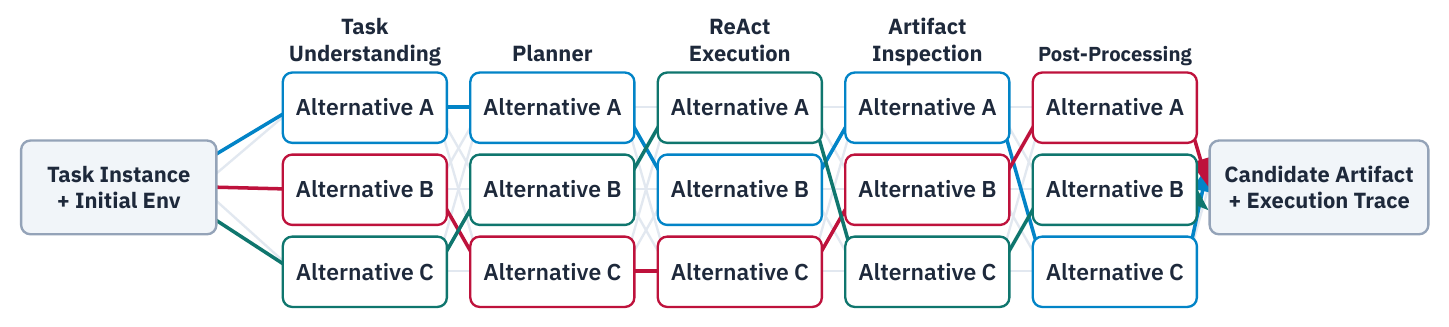}
    \caption{
    Persistent multi-alternative workflow used in Stage~2.
    A candidate path selects one prompt or behavior alternative from each
    active node, producing an artifact and execution trace. Alternatives
    persist across rounds, enabling non-destructive reuse and recombination.
    }
    \label{fig:workflow}
\end{figure*}

Figure~\ref{fig:workflow} shows the Stage~2 workflow representation.
The five active nodes are task understanding, planning, ReAct execution,
artifact inspection, and post-processing. A fresh workflow contains one seed
alternative at each node except ReAct execution, which contains two. The
alternative caps are \(2,3,5,2,2\), respectively. The baseline task-understanding
alternative and the disabled post-processing alternative are protected from
eviction.

Alternatives are weighted by their historical mean internal score, with an
uninformed prior of \(0.5\). Each newly proposed alternative receives one
forced exploration path in the next round; remaining paths use weighted
selection. When a node reaches its cap, the lowest-scoring non-protected
alternative is evicted. The evaluator remains frozen throughout this process.

\subsection{Gates, Budgets, and Termination}

\revise{Table~\ref{tab:appendix_gates} lists the gates and resource caps used in all standardized runs.}

\begin{table*}[t]
    \centering
    \small
    \setlength{\tabcolsep}{5pt}
    \begin{tabularx}{\textwidth}{llX}
        \toprule
        Stage
        & Component
        & Gate or budget \\
        \midrule
        1
        & Internal construction
        & Positive acceptance \(\geq 0.90\); adversarial rejection
          \(\geq 0.80\). \\

        1
        & Post-hoc \revise{admission}
        & External agreement \(\geq 0.85\) for admission to the standardized
          Stage~2 study. \\

        2
        & Workflow search
        & Five paths per round; candidate concurrency \(1\); at most
          \(15\) rounds. \\

        2
        & Candidate resources
        & Candidate wall-time limit \(300\) seconds; token budget \(250\)K;
          ReAct limit at most \(20\) by default and \(24\) at the framework
          cap. \\

        2
        & Strong executor
        & At most \(40\) turns and \(600\) seconds per candidate. \\

        \revise{2}
        & \revise{Round timeout}
        & \revise{One hour per round subprocess (Filesystem GLM-5.2 and
          PostgreSQL campaigns); a round that exceeds it ends the run, which is
          counted as unsolved.} \\

        3
        & Test-Suite Generator
        & Case budget \(24\); at most three label/coverage gate retries. \\

        3
        & Verifier Evolver
        & Initial proposal plus at most three repairs; at most two
          environment-repair restarts; \(900\)-second call timeout. \\

        3
        & Outer loop
        & At most three \revise{refinement rounds}; Stage~2 subprocess timeout
          \(21{,}600\) seconds. \\
        \bottomrule
    \end{tabularx}
    \caption{Standardized experimental gates and resource caps.}
    \label{tab:appendix_gates}
\end{table*}

The Stage~3 coordinator distinguishes semantic termination from infrastructure
failure. Its terminal states include exact reference match, semantic reference
convergence, evaluator stabilization on a repeated difference, \revise{refinement-round-cap}
exhaustion, evaluator-refinement failure, test-suite generation failure,
case-budget exhaustion, unrecovered evaluator-environment failure, Stage~2
infrastructure failure, and information-boundary violation. All three primary
Filesystem refinement runs terminate through evaluator stabilization on a
repeated, already-covered difference\revise{; for \texttt{structure\_mirror} this
happens after the excluded, artifact-triggered second round
(Appendix~\ref{app:structure_mirror})}.

\subsection{Environment Adapters and Observability}

\revise{Table~\ref{tab:appendix_observability} summarizes how each environment
materializes candidate states and what the evaluator can observe.}

\begin{table*}[t]
    \centering
    \scriptsize
    \setlength{\tabcolsep}{4pt}
    \begin{tabularx}{\textwidth}{lXXX}
        \toprule
        Environment
        & Materialized candidate
        & Evaluator-visible state
        & Principal limitation \\
        \midrule
        Filesystem
        & Isolated directory tree
        & Paths, file contents, directory structure, and supported metadata
        & Volatile backup paths and duplicate artifacts require conservative
          canonicalization. \\

        PostgreSQL
        & Fresh isolated database clone
        & Canonical schema, relation, and row snapshots
        & Some object classes and large relations are not fully represented in
          the internal snapshot. \\

        Playwright
        & Isolated browser execution
        & Tool transcript and selected browser-state signals
        & Some tasks lack independent factual ground truth or an observable
          post-condition. \\

        Notion
        & Duplicated workspace
        & Canonical page, block, database, and row snapshots
        & Linked views and serialization changes can obscure or alter the
          graded state. \\

        GitHub
        & Isolated repository
        & Repository contents and supported issue, pull-request, and history
          state
        & Tool-surface and \revise{verifier-environment} limitations block broad
          standardized coverage. \\
        \bottomrule
    \end{tabularx}
    \caption{
    Category-specific materialization and observation interfaces. Evaluator
    construction is meaningful only when task-relevant success conditions are
    represented in the permitted observation interface.
    }
    \label{tab:appendix_observability}
\end{table*}

\section{End-to-End Case Study: \texttt{structure\_mirror}}
\label{app:structure_mirror}

\subsection{Task and Initial \texorpdfstring{\revise{Evaluator}}{Evaluator}}

The task asks the agent to reproduce the directory structure of
\texttt{complex\_structure/} under
\texttt{complex\_structure\_mirror/} without copying file contents. Each
mirrored empty directory must contain a \texttt{placeholder.txt} file storing
its absolute path. Directories containing more than two direct files are
discarded, directory names containing digits receive a
\texttt{\_processed} suffix, and non-digit names are preserved.

Stage~1 synthesized five challenge states \revise{(Table~\ref{tab:structure_mirror_stage1_cases})}: one complete positive case, one
clean negative, and three adversarial near-misses. All five expected labels
agree with the external verifier. The initial evaluator \(E^{(0)}\) contains
five deterministic check families.

\begin{table}[t]
    \centering
    \scriptsize
    \setlength{\tabcolsep}{3pt}
    \begin{tabularx}{\columnwidth}{lclX}
        \toprule
        Case
        & Label
        & Type
        & Targeted requirement \\
        \midrule
        \texttt{correct\_full\_mirror}
        & \(\checkmark\)
        & Positive
        & Complete transformed mirror. \\

        \texttt{mirror\_dir\_missing}
        & \(\times\)
        & Clean negative
        & Required mirror root is absent. \\

        \texttt{no\_processed\_suffix}
        & \(\times\)
        & Adversarial
        & Digit-containing names are not transformed. \\

        \texttt{discarded\_dir\_present}
        & \(\times\)
        & Adversarial
        & A directory with more than two direct files is retained. \\

        \texttt{missing\_placeholder}
        & \(\times\)
        & Adversarial
        & A mirrored empty directory lacks its placeholder. \\
        \bottomrule
    \end{tabularx}
    \caption{Initial \texttt{structure\_mirror} challenge states.}
    \label{tab:structure_mirror_stage1_cases}
\end{table}

The corresponding checks verify that the mirror root exists, required
digit-bearing names are transformed, discarded directories are absent,
non-digit names are preserved, and placeholders exist. This evaluator achieves
perfect agreement on the generated Stage~1 suite, but it does not yet verify
that the final mirror contains no invalid extra structure.

\subsection{False-Positive Exposure}

Under \(E^{(0)}\), the weak executor reaches internal convergence at round
three. The retained state \revise{\(F_0^{-}\)} nevertheless fails external
verification, producing one of the three Filesystem \IntExt{\ok}{\no} states.
The strong campaign provides an internally and externally accepted reference
state \(F^{+}\).

\revise{\(E^{(0)}\) checks placeholders only in five listed leaf directories.
\(F_0^{-}\) omits \texttt{project/src/main/resources} and leaves three leaf
directories without a placeholder, none of which \(E^{(0)}\) checks. The five
Stage~1 challenge states do not expose this gap: their missing-placeholder
case omits every placeholder at once, which the five listed paths already
detect.}

\subsection{\texorpdfstring{\revise{One Refinement Round}}{One Refinement Round}}

\revise{Table~\ref{tab:structure_mirror_trajectory} summarizes the primary
refinement run.} In the first \revise{refinement round}, the strong--weak
difference is organized into empty-directory, missing-directory, and
missing-file clusters. The Test-Suite Generator converts these clusters into
six paired challenge states, and the Verifier Evolver repairs \(E^{(0)}\) in one
attempt\revise{, with a three-line patch that adds the three missing placeholder
paths to the placeholder check}. Cumulative regression verifies that
\(E^{(1)}\) accepts \(F^{+}\), rejects \revise{\(F_0^{-}\)}, and preserves all
earlier case verdicts\revise{ (11 cases in total)}.

\revise{Workflow evolution under \(E^{(1)}\) converges at round three on a state
\(F_1^{-}\) that the external verifier also accepts. Its canonical tree hash,
\texttt{80aeae4e...}, is the same as that of the externally accepted state of
the separate controlled run described below.}

\begin{table*}[t]
    \centering
    \scriptsize
    \setlength{\tabcolsep}{3.5pt}
    \begin{tabularx}{\textwidth}{clXXXl}
        \toprule
        Step
        & Evaluator
        & Newly exposed difference
        & New evaluation evidence
        & Promoted check
        & Weak-state outcome \\
        \midrule
        Initial
        & \(E^{(0)}\)
        & \revise{--}
        & Five initial Stage~1 cases
        & Presence, \revise{renaming, discard}, and placeholder checks \revise{(5 listed directories)}
        & \revise{\(F_0^{-}\)}: \IntExt{\ok}{\no} \\

        \revise{Round 1}
        & \(E^{(1)}\)
        & Empty-directory, missing-directory, and missing-file clusters
        & Six paired counterfactual cases
        & \revise{Three placeholder paths added; regression rejects \(F_0^{-}\)}
        & \revise{\(F_1^{-}\): \IntExt{\ok}{\ok}} \\

        \revise{Round 2 (excluded)}
        & \(E^{(2)}\)
        & \revise{Duplicate directory created by file synchronization}
        & \revise{Reproduce/remove pair}
        & \revise{Extra-structure absence check}
        & \revise{\IntExt{\no}{\no} (not counted)} \\
        \bottomrule
    \end{tabularx}
    \caption{
    Primary \texttt{structure\_mirror} evaluator lineage. \revise{One revision
    suffices: the weak state reached under \(E^{(1)}\) passes the external
    verifier. The second round was triggered by a file-synchronization artifact
    and is excluded from our results.}
    }
    \label{tab:structure_mirror_trajectory}
\end{table*}

\subsection{\texorpdfstring{\revise{An Artifact-Triggered Second Round}}{An Artifact-Triggered Second Round}}

\revise{The run did not stop at \(F_1^{-}\). Its second refinement round was
triggered by a single extra root-level directory, \texttt{complex\_structure~2},
in the working copy of \(F_1^{-}\). That directory is absent from the archived
\(F_1^{-}\) and from its canonicalization record: it was created by file
synchronization of the working directory after the state was archived.
Re-verifying the archived \(F_1^{-}\) after the fact, the external verifier and
\(E^{(0)}\), \(E^{(1)}\), and \(E^{(2)}\) all accept it. We therefore exclude the
second round from our results. The \(E^{(2)}\) it produced adds an absence check
learned from the artifact, and the weak state reached under \(E^{(2)}\) (score
\(0.8333\), rejected internally and externally) is reported only for
completeness.}

\subsection{\texorpdfstring{\revise{Separate Controlled Run}}{Separate Controlled Run}}

\revise{A separate controlled run from the same source version also reaches an
\(E^{(1)}\) and then performs a second round. Its round-2 difference is mostly
genuine: eleven extra \texttt{placeholder.txt} files, mostly in non-empty
directories, and one directory from the discarded subtree, together with a
synchronization-artifact directory \texttt{complex\_structure~3}. Its \(E^{(2)}\)
adds checks against this extra structure, and the weak executor then produces
an internally and externally accepted state with tree hash
\texttt{80aeae4e...}, the state the primary run reached after one round. A
development run that stopped after one round (Table~\ref{tab:filesystem_stage3_full})
confirms that extra structure is a real failure mode: its weak state, accepted
by its \(E^{(1)}\), contains eleven extra placeholder files and fails external
verification.}

\section{Complete Experimental Results}
\label{app:full_results}

\subsection{Filesystem Stage~1 Results}

\begin{table*}[p]
\centering\scriptsize
\setlength{\tabcolsep}{5pt}\renewcommand{\arraystretch}{0.92}
\begin{tabular}{@{}lcccc@{}}
\toprule
Task & Cases (P/N/A) & External agreement & Evaluator SHA & Refinement \\
\midrule
author\_folders & 4 (1/3/3) & 4/4 & \texttt{cadf0721} & -- \\
budget\_computation & 6 (1/5/4) & 6/6 & \texttt{3ffc73ef} & -- \\
code\_locating & 4 (2/2/2) & 4/4 & \texttt{af458c70} & -- \\
contact\_information & 6 (1/5/2) & 6/6 & \texttt{31564641} & -- \\
dataset\_comparison & 4 (1/3/3) & 4/4 & \texttt{71bb6f84} & -- \\
debugging & 4 (1/3/3) & 4/4 & \texttt{06c927a2} & E1 \\
dispute\_review & 4 (1/3/3) & 4/4 & \texttt{49314ccb} & -- \\
duplicate\_name & 4 (1/3/0) & 4/4 & \texttt{4a81584a} & -- \\
duplicates\_searching & 6 (1/5/4) & 6/6 & \texttt{74fc6149} & -- \\
english\_talent & 4 (1/3/0) & 4/4 & \texttt{0abee865} & -- \\
file\_arrangement & 6 (1/5/4) & 6/6 & \texttt{98a38fd4} & -- \\
file\_merging & 5 (1/4/0) & 5/5 & \texttt{d269b616} & -- \\
file\_splitting & 5 (1/4/0) & 5/5 & \texttt{d5127367} & -- \\
find\_math\_paper & 3 (1/2/2) & 3/3 & \texttt{2c95cf0b} & -- \\
gradebased\_score & 4 (1/3/0) & 4/4 & \texttt{2c94bcfc} & -- \\
individual\_comments & 4 (1/3/3) & 4/4 & \texttt{73c2e2ce} & -- \\
music\_report & 6 (2/4/3) & 6/6 & \texttt{8bedfe47} & -- \\
organize\_legacy\_papers & 4 (1/3/3) & 4/4 & \texttt{205a13b0} & -- \\
output\_analysis & 4 (2/2/2) & 4/4 & \texttt{607113fc} & -- \\
pattern\_matching & 5 (1/4/0) & 5/5 & \texttt{3cf52711} & -- \\
project\_management & 5 (1/4/3) & 5/5 & \texttt{4d721109} & -- \\
requirements\_completion & 4 (2/2/2) & 4/4 & \texttt{a1d5746a} & -- \\
requirements\_writing & 4 (1/3/3) & 4/4 & \texttt{21c6fb05} & -- \\
size\_classification & 5 (1/4/0) & 5/5 & \texttt{693581a4} & -- \\
solution\_tracing & 4 (1/3/3) & 4/4 & \texttt{65a1e2b6} & -- \\
structure\_analysis & 5 (1/4/0) & 5/5 & \texttt{f374f700} & -- \\
structure\_mirror & 5 (1/4/3) & 5/5 & \texttt{1b24f899} & \revise{E1} \\
time\_classification & 5 (1/4/3) & 5/5 & \texttt{b8412da2} & -- \\
timeline\_extraction & 6 (1/5/4) & 6/6 & \texttt{7ed448b2} & -- \\
uppercase & 6 (1/5/0) & 6/6 & \texttt{5935a8c6} & E1 \\
\midrule
\textbf{Total (30 \revise{evaluators})} & \textbf{141 (34/107/62)} & \textbf{141/141} & -- & 3 refined \\
\bottomrule
\end{tabular}
\caption{Complete per-task Filesystem Stage~1 results. P, N, and A denote positive, negative, and adversarial challenge states. External agreement is measured only on the generated challenge suite.}
\label{tab:filesystem_stage1_full}
\end{table*}

The primary Filesystem study constructs \revise{30/30 evaluators with}
141 total cases: 34 positive and 107 negative states, including 62 adversarial
near-misses. All 141 generated labels agree with the external verifier.
Per-task Verifier-Evolver iteration counts were not recorded consistently and
are therefore omitted.

\revise{Table~\ref{tab:filesystem_stage1_full} lists, for each task, the case counts,
external agreement, the hash prefix of the frozen evaluator, and whether the
evaluator was later refined in Stage~3.}

\subsection{Filesystem Stage~2 Results}

\begin{table*}[p]
\centering\scriptsize
\setlength{\tabcolsep}{4.5pt}\renewcommand{\arraystretch}{0.92}
\begin{tabular}{@{}lccccccc@{}}
\toprule
 & \multicolumn{2}{c}{GLM-5.2 (weak)} & \multicolumn{2}{c}{Claude CLI (strong)} & \multicolumn{3}{c}{GLM-5.2 loop} \\
\cmidrule(lr){2-3}\cmidrule(lr){4-5}\cmidrule(lr){6-8}
Task & Seed (Int/Ext) & Evolved (Int/Ext) & Seed (Int/Ext) & Evolved (Int/Ext) & Final round & Best score & Tokens (M) \\
\midrule
author\_folders & $\times$/$\times$ & $\times$/$\times$ & \checkmark/\checkmark & \checkmark/\checkmark & 14 & \revise{0.00} & 2.70 \\
budget\_computation & $\times$/$\times$ & $\times$/$\times$ & $\times$/$\times$ & \checkmark/\checkmark & 14 & 0.40 & 2.80 \\
code\_locating & \checkmark/\checkmark & \checkmark/\checkmark & \checkmark/\checkmark & \checkmark/\checkmark & 0 & 1.00 & 0.09 \\
contact\_information & $\times$/$\times$ & $\times$/$\times$ & \checkmark/\checkmark & \checkmark/\checkmark & 14 & 0.40 & 2.47 \\
dataset\_comparison & $\times$/$\times$ & \checkmark/\checkmark & \checkmark/\checkmark & \checkmark/\checkmark & 12 & 1.00 & 2.61 \\
debugging & \checkmark/$\times$ & \checkmark/$\times$ & \checkmark/\checkmark & \checkmark/\checkmark & 0 & 1.00 & 0.13 \\
dispute\_review & $\times$/$\times$ & \checkmark/\checkmark & $\times$/$\times$ & \checkmark/\checkmark & 2 & 1.00 & 1.08 \\
duplicate\_name & $\times$/$\times$ & \checkmark/\checkmark & \checkmark/\checkmark & \checkmark/\checkmark & 1 & 1.00 & 0.19 \\
duplicates\_searching & $\times$/$\times$ & \checkmark/\checkmark & \checkmark/\checkmark & \checkmark/\checkmark & 13 & 1.00 & 1.84 \\
english\_talent & \checkmark/\checkmark & \checkmark/\checkmark & \checkmark/\checkmark & \checkmark/\checkmark & 0 & 1.00 & 0.11 \\
file\_arrangement & $\times$/$\times$ & \checkmark/\checkmark & $\times$/\checkmark & \checkmark/\checkmark & 10 & 1.00 & 1.32 \\
file\_merging & $\times$/$\times$ & \checkmark/\checkmark & \checkmark/\checkmark & \checkmark/\checkmark & 9 & 1.00 & 1.36 \\
file\_splitting & $\times$/$\times$ & $\times$/$\times$ & \checkmark/\checkmark & \checkmark/\checkmark & 1\revise{$^\ddagger$} & 0.09 & 0.13 \\
find\_math\_paper & \checkmark/\checkmark & \checkmark/\checkmark & \checkmark/\checkmark & \checkmark/\checkmark & 0 & 1.00 & 0.08 \\
gradebased\_score & $\times$/$\times$ & $\times$/$\times$ & $\times$/\checkmark & \checkmark/\checkmark & 14 & \revise{0.00} & 2.91 \\
individual\_comments & \checkmark/\checkmark & \checkmark/\checkmark & \checkmark/\checkmark & \checkmark/\checkmark & 0 & 1.00 & 0.20 \\
music\_report & \checkmark/\checkmark & \checkmark/\checkmark & \checkmark/\checkmark & \checkmark/\checkmark & 0 & 1.00 & 0.12 \\
organize\_legacy\_papers & $\times$/$\times$ & $\times$/$\times$ & \checkmark/\checkmark & \checkmark/\checkmark & 14 & 0.40 & 2.86 \\
output\_analysis & $\times$/$\times$ & \checkmark/\checkmark & \checkmark/\checkmark & \checkmark/\checkmark & 4 & 1.00 & 0.78 \\
pattern\_matching & $\times$/$\times$ & $\times$/$\times$ & \checkmark/\checkmark & \checkmark/\checkmark & 14 & 0.33 & 1.86 \\
project\_management & $\times$/$\times$ & $\times$/$\times$ & \checkmark/\checkmark & \checkmark/\checkmark & 14 & 0.50 & 2.46 \\
requirements\_completion & $\times$/$\times$ & \checkmark/\checkmark & $\times$/\checkmark & \checkmark/\checkmark & 2 & 1.00 & 0.21 \\
requirements\_writing & $\times$/$\times$ & \checkmark/\checkmark & $\times$/$\times$ & \checkmark/\checkmark & 8 & 1.00 & 1.69 \\
size\_classification & $\times$/$\times$ & \checkmark/\checkmark & \checkmark/\checkmark & \checkmark/\checkmark & 3 & 1.00 & 0.30 \\
solution\_tracing & $\times$/$\times$ & \checkmark/\checkmark & \checkmark/\checkmark & \checkmark/\checkmark & 1 & 1.00 & 1.00 \\
structure\_analysis & $\times$/$\times$ & $\times$/$\times$ & \checkmark/\checkmark & \checkmark/\checkmark & 0\revise{$^\ddagger$} & \revise{0.00} & 0.10 \\
structure\_mirror & $\times$/$\times$ & \checkmark/$\times$ & \checkmark/\checkmark & \checkmark/\checkmark & 3 & 1.00 & 0.63 \\
time\_classification & $\times$/$\times$ & \checkmark/\checkmark & \checkmark/\checkmark & \checkmark/\checkmark & 6 & 1.00 & 0.98 \\
timeline\_extraction & \checkmark/\checkmark & \checkmark/\checkmark & \checkmark/\checkmark & \checkmark/\checkmark & 0 & 1.00 & 0.10 \\
uppercase & \checkmark/$\times$ & \checkmark/$\times$ & \checkmark/\checkmark & \checkmark/\checkmark & 0 & 1.00 & 0.06 \\
\midrule
\textbf{Total (of 30)} & \textbf{8/6} & \textbf{21/18} & \textbf{24/27} & \textbf{30/30} & -- & -- & \textbf{33.18} \\
\bottomrule
\end{tabular}
\caption{Complete per-task Filesystem Stage~2 outcomes. Each cell under Seed and Evolved reports internal/external acceptance. Seed is the internally selected best round-0 candidate before workflow evolution under the corresponding campaign's recorded path budget \revise{(five paths for GLM-5.2, three for Claude)}. \revise{\IntExt{\ok}{\no} and \IntExt{\no}{\ok}{} counts: GLM seed 2 and 0; GLM evolved 3 and 0; Claude seed 0 and 3; Claude evolved 0 and 0. Final round, best score, and tokens refer to the GLM-5.2 loop;} best score is 1.00 for converged tasks and otherwise the maximum per-round internal score. \revise{$^\ddagger$Run ended early when a round subprocess exceeded a one-hour timeout; counted as unsolved.}}
\label{tab:filesystem_stage2_full}
\end{table*}

\begin{table*}[t]
    \revisefloat
    \centering
    \scriptsize
    \setlength{\tabcolsep}{5pt}
    \begin{tabular}{@{}llrrrrrrr@{}}
\toprule
Executor & Verdict & Seed & 95\% CI & Evolved & 95\% CI & Gained & Lost & \(p\) \\
\midrule
GLM-5.2 & external & 6/30 & [0.10, 0.37] & 18/30 & [0.42, 0.75] & 12 & 0 & <0.001 \\
GLM-5.2 & internal & 8/30 & [0.14, 0.44] & 21/30 & [0.52, 0.83] & 13 & 0 & <0.001 \\
Claude & external & 27/30 & [0.74, 0.97] & 30/30 & [0.89, 1.00] & 3 & 0 & 0.25 \\
Claude & internal & 24/30 & [0.63, 0.90] & 30/30 & [0.89, 1.00] & 6 & 0 & 0.031 \\
\bottomrule
\end{tabular}

    \caption{Paired uncertainty of the Filesystem seed-to-evolved gains on the
    same 30 tasks: Wilson 95\% intervals and the exact two-sided McNemar test
    on tasks whose outcome changed.}
    \label{tab:fs_significance}
\end{table*}

\revise{Table~\ref{tab:filesystem_stage2_full} reports seed and evolved verdicts for
both executors, together with the final or convergence round, best internal
score, and executor tokens of the weak-executor loop. Two weak-executor runs,
\texttt{file\_splitting} and \texttt{structure\_analysis}, ended after three
and two rounds when a round subprocess exceeded a one-hour timeout; both are
counted as unsolved.} The audited
weak round-0 baseline is 8/30 internally and 6/30 externally, with two
false positives and no false negatives. The evolved weak campaign reaches
21/30 internally and 18/30 externally. The strong campaign improves
from 24/30 internal and 27/30 external seed successes to 30/30 on
both measures.

\subsection{\texorpdfstring{\revise{Held-Out Evaluation on Agent-Produced States}}{Held-Out Evaluation on Agent-Produced States}}
\label{app:heldout}

\revise{\paragraph{Protocol.}
For every best-of-round state retained by the two Filesystem Stage~2 campaigns,
we copy the snapshot, remove copy-collision duplicates (files named
\texttt{<name>~2} that are byte-identical to \texttt{<name>}, created by
file synchronization after the runs), re-execute the task's Stage~1 frozen
evaluator in the same sandbox used during the campaigns, and run the external
verifier once on the same copy (verdict: exit code~0). No model is called, and
nothing is fed back into any evaluator or workflow. Identical states (same task
and content hash) are counted once in the ``distinct'' rows.

\paragraph{Data.}
Of 320 retained snapshots, 258 are valid campaign states. We exclude 60 Claude
snapshots that lie beyond their task's final round and have no round record
(left over from an earlier run under the same campaign name) and two Claude
\texttt{file\_arrangement} states whose recorded internal verdict does not
reproduce, since they may not be the states that were scored. On the remaining
states, every re-executed internal verdict matches the recorded one, and all 87
previously recorded external verdicts reproduce. For 71 states the external
verifier exits with code~1 after reporting missing output files; these are
counted as failures, as in the round-0 baselines.

\paragraph{Results.}
Table~\ref{tab:heldout_full} reports both pools; Table~\ref{tab:heldout_per_task}
gives the per-task breakdown. The three false positives are the final GLM-5.2
states of \texttt{uppercase}, \texttt{debugging}, and \texttt{structure\_mirror}.
The six distinct false negatives are Claude states of \texttt{gradebased\_score}
(three), \texttt{file\_arrangement}, and \texttt{requirements\_completion}, and one
GLM-5.2 state of \texttt{requirements\_writing}. Scoring the Stage~3 evaluators on
the same pool is uninformative: after removing the states used as Stage~3
anchors, only two states of the three refined tasks remain.}

\begin{table*}[t]
    \revisefloat
    \centering
    \scriptsize
    \setlength{\tabcolsep}{4pt}
    \begin{tabular}{@{}lrrrrrrrrr@{}}
\toprule
 & & \multicolumn{4}{c}{Int/Ext} & & & & \\
\cmidrule(lr){3-6}
Executor & States & \ok/\ok & \ok/\no & \no/\ok & \no/\no & Agr. & Prec. & Rec. & Spec. \\
\midrule
\multicolumn{10}{@{}l}{\emph{All retained round states}} \\
\quad GLM-5.2 (all) & 203 & 18 & 3 & 1 & 181 & 98.0 & 85.7 & 94.7 & 98.4 \\
\quad GLM-5.2 (distinct) & 85 & 18 & 3 & 1 & 63 & 95.3 & 85.7 & 94.7 & 95.5 \\
\quad Claude (all) & 55 & 30 & 0 & 16 & 9 & 70.9 & 100.0 & 65.2 & 100.0 \\
\quad Claude (distinct) & 44 & 30 & 0 & 5 & 9 & 88.6 & 100.0 & 85.7 & 100.0 \\
\quad Both (all) & 258 & 48 & 3 & 17 & 190 & 92.2 & 94.1 & 73.8 & 98.4 \\
\quad Both (distinct) & 122 & 41 & 3 & 6 & 72 & 92.6 & 93.2 & 87.2 & 96.0 \\
\multicolumn{10}{@{}l}{\emph{Final states only}} \\
\quad GLM-5.2 (all) & 30 & 18 & 3 & 0 & 9 & 90.0 & 85.7 & 100.0 & 75.0 \\
\quad Claude (all) & 30 & 30 & 0 & 0 & 0 & 100.0 & 100.0 & 100.0 & -- \\
\quad Both (all) & 60 & 48 & 3 & 0 & 9 & 95.0 & 94.1 & 100.0 & 75.0 \\
\quad Both (distinct) & 53 & 41 & 3 & 0 & 9 & 94.3 & 93.2 & 100.0 & 75.0 \\
\bottomrule
\end{tabular}

    \caption{Held-out evaluation of the Stage~1 frozen evaluators on
    agent-produced Filesystem states, by executor and pool. ``All'' counts every
    retained state; ``distinct'' counts identical states once. Metrics in \%.}
    \label{tab:heldout_full}
\end{table*}

\begin{table*}[t]
    \revisefloat
    \centering
    \scriptsize
    \setlength{\tabcolsep}{4pt}
    \begin{tabular}{@{}lrrrrr@{}}
\toprule
Task & States & \ok/\ok & \ok/\no & \no/\ok & \no/\no \\
\midrule
author\_folders & 2 & 1 & 0 & 0 & 1 \\
budget\_computation & 4 & 1 & 0 & 0 & 3 \\
code\_locating & 2 & 2 & 0 & 0 & 0 \\
contact\_information & 4 & 1 & 0 & 0 & 3 \\
dataset\_comparison & 10 & 2 & 0 & 0 & 8 \\
debugging & 2 & 1 & 1 & 0 & 0 \\
dispute\_review & 4 & 2 & 0 & 0 & 2 \\
duplicate\_name & 3 & 2 & 0 & 0 & 1 \\
duplicates\_searching & 6 & 1 & 0 & 0 & 5 \\
english\_talent & 1 & 1 & 0 & 0 & 0 \\
file\_arrangement & 8 & 2 & 0 & 1 & 5 \\
file\_merging & 7 & 2 & 0 & 0 & 5 \\
file\_splitting & 3 & 1 & 0 & 0 & 2 \\
find\_math\_paper & 1 & 1 & 0 & 0 & 0 \\
gradebased\_score & 5 & 1 & 0 & 3 & 1 \\
\bottomrule
\end{tabular}\hfill
\begin{tabular}{@{}lrrrrr@{}}
\toprule
Task & States & \ok/\ok & \ok/\no & \no/\ok & \no/\no \\
\midrule
individual\_comments & 1 & 1 & 0 & 0 & 0 \\
music\_report & 1 & 1 & 0 & 0 & 0 \\
organize\_legacy\_papers & 7 & 1 & 0 & 0 & 6 \\
output\_analysis & 4 & 2 & 0 & 0 & 2 \\
pattern\_matching & 4 & 1 & 0 & 0 & 3 \\
project\_management & 3 & 1 & 0 & 0 & 2 \\
requirements\_completion & 6 & 2 & 0 & 1 & 3 \\
requirements\_writing & 14 & 2 & 0 & 1 & 11 \\
size\_classification & 4 & 1 & 0 & 0 & 3 \\
solution\_tracing & 3 & 2 & 0 & 0 & 1 \\
structure\_analysis & 2 & 1 & 0 & 0 & 1 \\
structure\_mirror & 4 & 1 & 1 & 0 & 2 \\
time\_classification & 4 & 2 & 0 & 0 & 2 \\
timeline\_extraction & 1 & 1 & 0 & 0 & 0 \\
uppercase & 2 & 1 & 1 & 0 & 0 \\
\bottomrule
\end{tabular}

    \caption{Held-out evaluation per task: distinct retained states from both
    executors, by internal/external verdict.}
    \label{tab:heldout_per_task}
\end{table*}

\subsection{Filesystem Stage~3 Lineage}

\begin{table}[t]
    \centering
    \scriptsize
    \setlength{\tabcolsep}{2.2pt}
    \begin{tabular}{@{}p{0.19\columnwidth}
                    p{0.30\columnwidth}
                    p{0.18\columnwidth}
                    p{0.24\columnwidth}@{}}
        \toprule
        Task
        & Diff-exposed constraint
        & Revision
        & Outcome \\
        \midrule
        \texttt{uppercase}
        & Exact text and terminal-newline fidelity
        & \(E^{(0)}\!\rightarrow E^{(1)}\) (1)
        & FP removed; final
          \IntExt{\no}{\no} \\

        \texttt{debugging}
        & Presence of the required fixed code pattern
        & \(E^{(0)}\!\rightarrow E^{(1)}\) (3)
        & FP removed; final
          \IntExt{\no}{\no} \\

        \texttt{structure\_\allowbreak mirror}
        & \revise{Missing nested directories and leaf placeholders}
        & \(E^{(0)}\!\rightarrow E^{(1)}\) \revise{(1)}
        & FP removed; \revise{next state
          \IntExt{\ok}{\ok}} \\
        \bottomrule
    \end{tabular}
    \caption{Execution-state refinement on the three Filesystem false
    positives. Parentheses give Verifier-Evolver calls for each accepted
    revision. \revise{On \texttt{uppercase} and \texttt{debugging}, the final
    weak states are \IntExt{\no}{\no}; on \texttt{structure\_mirror}, the
    weak state reached under \(E^{(1)}\) passes both. A second
    \texttt{structure\_mirror} round, triggered by a file-synchronization
    artifact, is excluded (Appendix~\ref{app:structure_mirror}).}}
    \label{tab:rq3_refinement}
\end{table}

\begin{table*}[t]
\centering\scriptsize
\setlength{\tabcolsep}{4.5pt}\renewcommand{\arraystretch}{1.0}
\begin{tabularx}{\textwidth}{@{}llcclX@{}}
\toprule
Task & Provenance class & \revise{Evaluators} & Final Int./Ext. & Terminal & Key result \\
\midrule
uppercase & \revise{Primary refinement run} & E0$\to$E1 & $\times$/$\times$ & stabilized (r1) & Original false positive removed (exact-match check replaces \texttt{.strip()} tolerance). \\
debugging & \revise{Primary refinement run} & E0$\to$E1 & $\times$/$\times$ & stabilized (r1) & Original false positive removed (require-fix-present check via paired contrast). \\
structure\_mirror & \revise{Primary refinement run} & E0$\to$E1\revise{ (E2 excluded)} & \revise{\checkmark/\checkmark\ after r1} & stabilized (r2) & \revise{Original false positive removed; the weak state under E1 is externally valid (hash \texttt{80aeae4e}). Round 2 was triggered by a file-synchronization artifact and is excluded.} \\
structure\_mirror & FP exposure (development case study) & E0$\to$E1 & \checkmark/$\times$ & semantic convergence & \revise{Extra-structure blind spot: E1 accepts a state with eleven extra placeholder files, which fails external verification.} \\
structure\_mirror & \revise{Separate controlled run (external pass)} & E0$\to$E1$\to$E2 & \checkmark/\checkmark & stabilized (r2) & \revise{Round 2 targets genuine extra structure (plus a synchronization artifact); final state hash \texttt{80aeae4e}, identical to the primary run's round-1 state.} \\
\bottomrule
\end{tabularx}
\caption{Primary and case-study Filesystem Stage~3 trajectories. \revise{Primary refinement runs, the false-positive-exposure run, and the separate controlled run are separated by provenance class. The final weak states of the \texttt{uppercase} and \texttt{debugging} primary runs are rejected internally and externally (calibrated true negatives); the \texttt{structure\_mirror} primary run's state after round 1 passes both. The controlled run is not pooled with the primary runs.}}
\label{tab:filesystem_stage3_full}
\end{table*}

The complete Stage~3 ledger records 42 primary, development, diagnostic, and
case-study trajectories. Primary claims use only the three \revise{primary refinement}
runs and the explicitly separated \texttt{structure\_mirror} exposure and
\revise{controlled} runs.

\subsection{Cross-Environment Results}
\label{app:category_results}

\revise{Table~\ref{tab:cross_environment_results} summarizes the cross-environment evidence by stage.}

\begin{table*}[t]
    \centering
    \scriptsize
    \setlength{\tabcolsep}{4pt}
    \begin{tabularx}{\textwidth}{lXXX}
        \toprule
        Environment
        & Stage~1 evidence
        & Stage~2 evidence
        & Stage~3 disposition \\
        \midrule

        PostgreSQL
        & 18/21 evaluators successfully constructed over 120 calibration states
          (22 positive / 98 negative), with 115/120 internal--external
          agreement; 15/21 meet the Stage~2 \revise{admission threshold}
        & 15 matched pairs run to the configured limits; both executors solve
          three tasks. The strong executor succeeds at round~0 on all three,
          while the weak executor succeeds after rounds 11, 2, and 7;
          \revise{neither converges on the remaining 12 tasks (8 weak-executor
          runs end at the one-hour round timeout)}
        & No agent-produced
          \IntExt{\ok}{\no} state across the 30
          Stage~2 conditions; refinement not triggered \\

        Playwright
        & Four evaluators over 88 generated cases; two correctness evaluators
          and two format-contract-only evaluators
        & Two valid pairs: both executors pass \texttt{extraction\_table};
          both fail \texttt{r1\_arxiv}
        & No eligible false positive; two tasks formally excluded \\

        Notion
        & Two loop-native evaluators calibrate at agreement \(1.0\);
          one additional evaluator is \revise{blocked by the observation interface}
        & One historical matched pair on \texttt{remove\_osaka}: strong
          condition \IntExt{\ok}{\no}, weak
          condition \IntExt{\no}{\no}
        & No completed eligible refinement pair; outer refinement not run \\

        GitHub
        & One full calibration at \(0.857\); additional Stage~1 \revise{evaluators} remain
          incomplete, divergent, or \revise{blocked by the observation interface}
        & Corrected canonical-snapshot scoring yields one provenance-valid
          strong-executor \(F^{+}\) on \texttt{find\_salient\_file};
          a corrected weak-executor matched run \revise{was not completed}
        & No corrected
          \IntExt{\ok}{\no} pair; refinement not
          triggered \\
        \bottomrule
    \end{tabularx}
    \caption{
    Cross-environment evidence. PostgreSQL provides the broadest
    cross-environment Stage~1/Stage~2 study; Playwright, Notion, and GitHub
    remain smaller or incomplete studies. These results are not pooled into the
    Filesystem task-level aggregate.
    }
    \label{tab:cross_environment_results}
\end{table*}

\paragraph{PostgreSQL.}
Across all 21 standard PostgreSQL tasks, VeriEvolve successfully constructs
evaluators for 18 tasks. Their 120 post-hoc calibration states contain 22
external-positive and 98 external-negative cases; the frozen internal
evaluators agree with the external verifier on 115/120 states. Fifteen
evaluators satisfy the per-task \(0.85\) external-agreement threshold and enter
Stage~2, three successfully constructed evaluators remain below threshold, and
three tasks are \revise{blocked by limitations of the observation
interface}. External calibration is performed only after evaluator construction
and is not routed back into evaluator generation or repair.

All 15 eligible tasks receive matched strong- and weak-executor Stage~2 runs.
Both conditions ultimately solve
\texttt{employee\_hierarchy\_management},
\texttt{customer\_data\_migration}, and
\texttt{sales\_and\_music\_charts}. The strong executor succeeds at round~0 on
all three, whereas the weak executor reaches externally verified success after
11, 2, and 7 workflow-evolution rounds, respectively. The weak/strong executor
token ratios on these tasks are \(32.2\times\), \(5.2\times\), and
\(12.8\times\). On the remaining 12 eligible tasks, \revise{neither condition reaches internal
convergence. All 12 strong-executor runs use the full 15 rounds; of the 12
weak-executor runs, 4 use all 15 rounds and 8 end earlier, after 4 to 12
rounds, when a round exceeds the one-hour round timeout.}

Across the resulting 30 Stage~2 conditions, every internally accepted final
state is also externally accepted: six conditions are
\IntExt{\ok}{\ok}, while the remaining 24
conditions do not reach internal acceptance. No agent-produced
\IntExt{\ok}{\no} state therefore occurs, so
PostgreSQL does not trigger execution-state refinement. This separates the
existence of evaluator imperfections from their exposure by ordinary agent
trajectories.

\paragraph{Playwright.}
The \texttt{extraction\_table} pair reaches internal and benchmark-external
acceptance under an oracle that primarily validates output format and row
count; this should not be interpreted as a complete semantic-correctness
guarantee. The \texttt{r1\_arxiv} pair remains at internal and external
failure under a correctness-calibrated evaluator. The remaining two tasks are
excluded because either independent ground truth or an observable success
state is unavailable.

\paragraph{Notion.}
The two calibrated Stage~1 \revise{evaluators} use small suites and are therefore
indicative rather than category-wide evidence. The
\texttt{remove\_osaka} historical matched pair exposes an evaluator
divergence, but its evaluator is below the \(0.85\) \revise{admission threshold}.
Controlled serialization analysis measures agreement \(0.75\) under the
historical schema and \(0.50\) after a live-state serialization change. The
latter decrease is an engineering regression layered on top of the underlying
semantic divergence.

\paragraph{GitHub.}
The GitHub study remains an incomplete feasibility campaign. A \revise{subsequent}
audit found that the historical Stage~2 zero scores were produced by evaluating
the live backend inside a short sandbox timeout; those runs are therefore not
used as agent-performance evidence. After replacing live-backend evaluation
with canonical-snapshot scoring, a strong-executor run on
\texttt{find\_salient\_file} converges at round~0 and passes post-hoc external
verification. The corresponding corrected weak-executor matched run \revise{was not
completed}, so we make no GitHub matched-pair or Stage~3 claim.

\section{Additional Analyses and Ablations}
\label{app:ablations}

\subsection{Workflow Representation and Evolution}
\label{app:workflow_ablation}

Table~\ref{tab:workflow_ablation} compares the primary campaigns with an
earlier React-only Claude campaign. The historical run evolved only the
execution prompt and improved external success from 17/30 to 30/30.
The final five-node workflow begins much higher for Claude, at 27/30
external seed success, and evolution closes the final three tasks. For
GLM-5.2, the same final workflow improves from 6/30 to 18/30.

\begin{table*}[t]
    \centering
    \small
    \setlength{\tabcolsep}{5pt}
    \begin{tabularx}{\textwidth}{llXrr}
        \toprule
        Campaign
        & Executor
        & Workflow and evolver
        & Seed Ext.
        & Final Ext. \\
        \midrule
        Historical
        & Claude
        & Task understanding plus ReAct execution; only the ReAct prompt
          evolved by an interactive Claude Code outer loop
        & 17/30
        & 30/30 \\

        Primary
        & Claude
        & Five-node persistent multi-alternative workflow;
          headless Workflow Evolver
        & 27/30
        & 30/30 \\

        Primary
        & GLM-5.2
        & Same five-node representation and task-specific evaluators
        & 6/30
        & 18/30 \\
        \bottomrule
    \end{tabularx}
    \caption{
    Workflow representation and evolution results. The historical and primary
    Claude campaigns differ in representation, seed, source version, and
    evolver invocation, so the comparison is supporting rather than a
    controlled single-variable ablation.
    }
    \label{tab:workflow_ablation}
\end{table*}

The comparison is consistent with complementary contributions from the
workflow scaffold and evaluator-guided evolution. A richer scaffold places
more strong-model capability in the seed, while evolution provides a larger
marginal gain for the weaker executor. Because several experimental components
changed between the two Claude campaigns, the 17/30 versus 27/30
difference should not be attributed solely to the five-node representation.

\subsection{Weak-Executor Context and Model Tier}

Removing context truncation substantially changes the weak-executor result.
The truncated GLM-5.2 campaign converges internally on 7/18 completed
tasks, whereas the untruncated campaign converges on 21/30. Twelve tasks
previously classified as stalled become internally converged under full
history. The comparison is not denominator-matched, but the task-level audit
supports context truncation as a major source of the earlier failure mode.

Under identical task-specific evaluators, GLM-4.5 reaches 9/30 internal
and 7/30 external successes, compared with 21/30 and 18/30 for
GLM-5.2. A separate history-accumulating replication reaches 14/30
internally and 12/30 externally, with the same \(85.7\%\) internal
precision as the primary weak campaign.

\subsection{Paired Contrasts and Localized Evaluator Repair}

On \texttt{debugging}, unaided evaluator regeneration fails in four attempts.
Providing an API-visible positive--negative contrast and constraining the
Verifier Evolver to a localized patch yields the required
``fixed-code-present'' check in two to three attempts. The corrected evaluator
then assigns full internal fitness to a valid executor state. This analysis
supports the final Stage~3 design choice to expose factual paired contrasts
while requiring minimal, cumulative-regression-checked evaluator changes.

\subsection{Stagnation and Redundant Refinement}

A cap-three replay on \texttt{uppercase} produces identical later weak states
and content-identical evaluator revisions. Two of the three additional
\revise{refinement rounds} therefore add no new evaluator capability. The final stagnation
gate detects that the remaining state difference is already covered and
terminates all three primary \revise{refinement} runs without generating redundant cases.

\subsection{\texorpdfstring{\revise{Strong-Executor Reliability}}{Strong-Executor Reliability}}
\label{app:strong_executor}

\revise{Stage~3 needs a reference state \(F^{+}\) that the strong executor
reaches and that is accepted internally and externally.
Table~\ref{tab:strong_executor} counts, for each environment, how often the
strong executor (Claude) provides one in the paper's runs. It does on all 30
Filesystem tasks but on only 3 of the 15 admitted PostgreSQL tasks, so
refinement could not run on the remaining 12 even if a weak-executor false
positive had appeared there. Of the 36 strong-executor states accepted by their
internal evaluator, one (Notion \texttt{remove\_osaka}) is rejected
externally: trusted without the external check, it would have been a wrong
reference. In our experiments every \(F^{+}\) used for refinement was
externally accepted, so no such error entered Stage~3; how a wrong reference
would affect the revised evaluator is untested.}

\begin{table}[t]
    \revisefloat
    \centering
    \scriptsize
    \setlength{\tabcolsep}{2.5pt}
    \begin{tabular}{@{}lrrrr@{}}
\toprule
Environment & Tasks & Usable \(F^{+}\) & \IntExt{\ok}{\no} & Int\,\(\times\) \\
\midrule
Filesystem & 30 & 30 & 0 & 0 \\
PostgreSQL & 15 & 3 & 0 & 12 \\
Playwright & 2 & 1 & 0 & 1 \\
Notion & 1 & 0 & 1 & 0 \\
GitHub & 1 & 1 & 0 & 0 \\
\midrule
Total & 49 & 35 & 1 & 13 \\
\bottomrule
\end{tabular}

    \caption{Strong-executor (Claude) outcomes in the paper's runs, one final
    run per task. A usable reference \(F^{+}\) is a state accepted both
    internally and externally (\IntExt{\ok}{\ok}); Int\,\(\times\) counts
    states not accepted internally.}
    \label{tab:strong_executor}
\end{table}

\subsection{\texorpdfstring{\revise{A Verifier-Free Refinement Trigger}}{A Verifier-Free Refinement Trigger}}
\label{app:trigger}

\revise{In our experiments, Stage~3 is applied to the internally accepted states
that the external verifier rejects. We test retrospectively whether executor
disagreement could replace that signal. For each of the 21 GLM-5.2 final states
that its frozen evaluator accepts, we compare it with the strong executor's
accepted final state on the same task, after the same canonicalization as in
Stage~3, and flag the task under several rules on the resulting difference.
External verdicts are used only to score the rules
(Table~\ref{tab:trigger}).

Flagging any difference catches all three false positives: eight accepted weak
states are identical to the strong state and are correctly left alone. However,
ten of the 18 correct states also differ, mostly in file content, so only 3 of
the 13 flagged tasks are real false positives. Narrower rules miss some false
positives, because the three differ in different ways: \texttt{uppercase} only
in trailing newlines, \texttt{debugging} in file content, and
\texttt{structure\_mirror} in missing structure. Such a trigger is therefore
usable only if Stage~3 can dismiss harmless differences. Its
instruction-grounding filter and regression gate are designed to do this, but
they were not exercised on these false alarms, so whether they would decline
them or over-tighten the evaluator is untested. The analysis also requires a
strong-executor state and uses only three positive cases.}

\begin{table}[t]
    \revisefloat
    \centering
    \scriptsize
    \setlength{\tabcolsep}{2pt}
    \resizebox{\columnwidth}{!}{\begin{tabular}{@{}lrrr@{}}
\toprule
Trigger rule (weak vs.\ strong accepted state) & Flagged & FPs caught & Correct states flagged \\
\midrule
Any difference & 13/21 & 3/3 & 10/18 \\
Structural difference (missing/extra/empty) & 4/21 & 1/3 & 3/18 \\
Missing structure only (strong has it, weak lacks it) & 2/21 & 1/3 & 1/18 \\
Content difference (excluding newline-only) & 10/21 & 1/3 & 9/18 \\
Any difference, newline-only differences ignored & 12/21 & 2/3 & 10/18 \\
\bottomrule
\end{tabular}
}
    \caption{Retrospective verifier-free triggers on the 21 internally accepted
    GLM-5.2 final states (3 false positives, 18 correct), each compared with the
    strong executor's accepted state on the same task.}
    \label{tab:trigger}
\end{table}

\subsection{Environment-Specific Diagnostics}

\revise{Table~\ref{tab:environment_diagnostics} lists diagnostic findings from the
cross-environment studies that bear on how their results should be read.}

\begin{table*}[t]
    \centering
    \scriptsize
    \setlength{\tabcolsep}{4pt}
    \begin{tabularx}{\textwidth}{lXX}
        \toprule
        Analysis
        & Finding
        & Interpretation \\
        \midrule

        PostgreSQL numeric type
        & An evaluator rejects valid \texttt{Decimal} values because it tests
          only \texttt{int} and \texttt{float}; the external verifier accepts
          the state
        & Evaluator type mismatch rather than executor failure; corrected by a
          type-equivalence update before the final campaign. \\

        PostgreSQL Stage~2 \revise{non-convergence}
        & \revise{Neither executor converges} on 12/15 Stage~2-eligible
          tasks despite evaluators that cleared the \revise{admission threshold}
        & A usable fitness signal does not guarantee task completion; workflow
          search, executor capability, and interaction budget remain limiting. \\

        Playwright observability
        & Turnstile exposes no stable transcript-visible success signal;
          another task lacks independently grounded factual truth
        & Formal exclusions rather than task failures. \\

        Notion serialization
        & Agreement falls from \(0.75\) to \(0.50\) when people and date fields
          change serialization
        & A real evaluator disagreement compounded by backward-incompatible
          state representation. \\

        GitHub scoring \revise{interface}
        & Historical in-loop evaluator calls queried the live backend inside a
          short sandbox timeout, producing vacuous zero scores; canonical
          snapshot scoring restores valid internal evaluation
        & Historical non-convergence is excluded from performance claims;
          corrected strong-executor execution demonstrates feasibility, while
          the standardized matched pair remains incomplete. \\
        \bottomrule
    \end{tabularx}
    \caption{Environment-specific diagnostic findings.}
    \label{tab:environment_diagnostics}
\end{table*}

\section{Compute, Cost, and Runtime}
\label{app:compute}

Table~\ref{tab:compute_accounting} reports primary experimental accounting.
Stage~1 was performed through interactive sessions across multiple context
windows, so its task-level token and runtime costs were not recorded and are
not estimated. Diagnostic and development campaigns are excluded from the
primary totals.

\begin{table*}[t]
    \centering
    \small
    \setlength{\tabcolsep}{3.5pt}
    \begin{tabularx}{\textwidth}{lrrrrX}
        \toprule
        Experiment
        & Tasks/runs
        & Executor tokens
        & Evolver cost
        & Recorded wall
        & Qualification \\
        \midrule

        Filesystem Stage~1
        & 30
        & --
        & --
        & --
        & Interactive accounting not recorded. \\

        Filesystem Stage~2, Claude
        & 30
        & 575{,}739
        & \$1.74
        & \revise{5.3 task-hours}
        & Primary strong campaign\revise{; executor tokens exclude prompt-cache reads}. \\

        Filesystem Stage~2, GLM-5.2
        & 30
        & 33{,}176{,}550
        & \$31.24
        & 43.5 task-hours
        & Primary weak campaign; tasks executed across parallel lanes. \\

        Filesystem Stage~3
        & 3
        & 8{,}100{,}868
        & \$3.49
        & \(\approx 2.5\) h actual
        & Primary \revise{refinement runs}; TSG and VE additionally use
          37{,}031 and 27{,}235 Sonnet tokens. \\

        PostgreSQL matched pairs
        & 30 conditions
        & \(\approx 46.9\)M
        & \$41.88
        & \(\approx 71\) h loop wall
        & Full 15-pair Stage~2 campaign over all eligible tasks. \\

        Playwright matched pairs
        & 4 conditions
        & 1{,}062{,}274
        & \$1.71
        & 94 min
        & Two valid matched pairs. \\

        Notion historical pair
        & 2 conditions
        & \(\approx 110\)K
        & \$0.50
        & --
        & Historical diagnostic pair; not pooled with standardized campaigns. \\

        GitHub corrected \revise{strong-executor} pilot
        & 1 condition
        & 193{,}784
        & --
        & --
        & Provenance-valid strong \(F^{+}\); corrected weak matched run
          incomplete. \\
        \bottomrule
    \end{tabularx}
    \caption{
    Primary and qualified experimental compute accounting. Executor tokens and
    evolver cost are reported separately. Missing measurements are left
    unestimated; historical or incomplete pilot rows are explicitly labeled.
    }
    \label{tab:compute_accounting}
\end{table*}

The weak Filesystem campaign \revise{records} approximately \(58\times\) the executor
tokens of the strong campaign\revise{; Claude executor counts exclude prompt-cache
reads, so this ratio is indicative rather than exact}. Stage~3 is also expensive because each accepted
evaluator revision triggers another five-path, up-to-15-round weak-executor
search. The expanded PostgreSQL campaign further illustrates that evaluator
availability does not make search inexpensive: running both conditions over
the 15 eligible tasks requires approximately \(46.9\)M executor tokens, while
most tasks still \revise{end without internal convergence}. These costs motivate
selective execution-state refinement rather than universal Stage~3 sweeps over
all benchmark tasks.

The PostgreSQL campaign was checkpointed across transient provider failures;
completed rounds were resumed without recomputation or changes to scientific
settings. Similarly, the three Filesystem Stage~3 runs were initially launched
in parallel, but concurrent Workflow-Evolver calls caused correlated transient
provider failures. They were resumed sequentially from completed checkpoints
without changing scientific settings. Parallelism therefore reduced actual
wall time while introducing provider-contention overhead.

\section{Oracle Isolation, Provenance, and Reproducibility}
\label{app:integrity}

\subsection{External-Verifier Information Boundary}

The external MCPMark verifier is invoked only after an internal evaluator or
workflow trajectory has been frozen. Its verdict is never used for workflow
ranking, alternative credit, evaluator repair, convergence, or termination.
\revise{Outside the loop, its verdicts are used for three decisions: admitting an
evaluator to Stage~2 (agreement \(\geq 0.85\) on the evaluator's challenge
suite), selecting the internally accepted but externally rejected states that
Stage~3 refines, and confirming that each reference state \(F^{+}\) passes.}
The strength of this guarantee differs across experimental phases and is
reported explicitly in Table~\ref{tab:oracle_isolation}.

\begin{table*}[t]
    \centering
    \scriptsize
    \setlength{\tabcolsep}{3.5pt}
    \begin{tabularx}{\textwidth}{lXXX}
        \toprule
        Phase
        & Technically enforced controls
        & Procedural restrictions
        & Audit evidence \\
        \midrule
        Interactive Stage~1
        & No hard filesystem-level isolation is proven by retained artifacts
        & Workspace policy prohibits reading verifier source; external
          calibration occurs after evaluator construction
        & External fields appear only in post-construction calibration
          artifacts. Claims for this phase are procedural, not technical. \\

        Automated Stage~2
        & Workflow-Evolver inputs are constructed only from internal scores,
          failed checks, traces, and alternative statistics; executors are
          restricted to isolated task environments
        & Prompts prohibit reference to external verifier logic
        & Source/input hashes, routing policies, and run manifests are
          retained. \\

        Stage~3 TSG and VE
        & Packet-only temporary working directory outside the repository;
          shell, search, write, web, and task tools disabled; packet allowlists
          and forbidden-token scans; tool-access logging and run-aborting
          outside-access checks
        & Read access is additionally restricted by the packet contract
        & Per-round packet hashes and access-decision records; primary runs
          contain no information-boundary violation. \\

        Post-hoc verification
        & Separate subprocess with provider credentials stripped; bounded
          output capture; invoked only after internal termination
        & External result cannot be routed back into a model role
        & External records contain invocation status and return code and are
          stored outside model packets. \\
        \bottomrule
    \end{tabularx}
    \caption{
    External-verifier isolation by phase. Initial interactive Stage~1 relies on
    a documented procedural boundary; Stage~2 and especially Stage~3 add
    technically enforced input and tool restrictions.
    }
    \label{tab:oracle_isolation}
\end{table*}

The effective information flow is:

{\small
\[
\begin{aligned}
x,\mathcal{O}(s_0)
&\rightarrow
\mathrm{TSG}
\rightarrow
\mathcal{C}^{(0)}
\rightarrow
\mathrm{VE}
\rightarrow
E^{(0)},\\
E^{(k)}
&\rightarrow
\text{WorkflowSearch}
\rightarrow
F_k^{-}\\
&\qquad\rightarrow
\revise{\Delta^{(k)}}
\rightarrow
\mathcal{C}^{(k+1)},E^{(k+1)},\\
\text{Frozen trajectory}
&\rightarrow
\text{external verification}.
\end{aligned}
\]
}

There is no return edge from the final line to any generation, scoring,
\revise{repair}, ranking, or stopping component\revise{; the only uses of external
verdicts are the three selection decisions above, which happen between runs.}

\subsection{Stage~3 Packet Isolation}

Each Stage~3 model call receives a small allowlisted packet containing the task
instruction, current evaluator, sanitized challenge cases, factual state-diff
information, and an output contract. State paths, external verdict fields, and
other non-allowlisted values are removed before packet creation. Every packet
is scanned for prohibited tokens, including references to the external
verifier and external verdict fields.

The model runs from a temporary directory outside the workbench. Most
filesystem and process tools are disabled; read operations are logged. An
attempt to access material outside the approved packet is classified as an
\texttt{information\_boundary\_violation} and aborts the run. No such
violation occurs in the primary Stage~3 campaign.

\subsection{Provenance and Immutable Artifacts}

Each reported run retains its source commit, strategy identifier, evaluator
versions and hashes, workflow-state hashes, per-round telemetry, and terminal
classification. Stage~3 state \revise{archives} preserve canonical tree hashes and
archive hashes for all retained weak and reference states. Cumulative
regression records whether the candidate evaluator accepts \(F^{+}\), rejects
all prior false-positive weak states, preserves previous case verdicts, and
handles newly added cases.

Independent reproduction provides additional evidence:

\begin{itemize}
    \item The \texttt{uppercase} \(E^{(1)}\) evaluator is byte-identical across
    three independent Verifier-Evolver searches.
    \item Canonical weak-state tree hashes recur across independent runs for
    \texttt{uppercase}, \texttt{debugging}, and
    \texttt{structure\_mirror}.
    \item The externally accepted
    \texttt{structure\_mirror} state hash
    \texttt{80aeae4e...} is \revise{reached both by the primary run after one
    refinement round and by the separate controlled run}.
\end{itemize}

\subsection{Run Classification and Intervention Policy}

Infrastructure failures are separated from semantic task outcomes.
Provisioning errors, provider aborts, source mismatches, and unobservable
success states are not assigned a semantic score of zero. The primary
campaigns permit deterministic materialization and harness repair,
checkpoint-resume after transient provider failures, and bounded model repair
within the predeclared budget. They do not permit task-specific manual
evaluator logic, external-verifier feedback inside a loop, or unreported
changes to scientific settings.

Development and diagnostic runs are retained for mechanism analysis but are
not pooled with primary outcomes. This distinction is especially important
for the Stage~3 development lineage, the pre-fix GitHub scoring runs, the
Notion serialization diagnostics, and the pre-provenance Playwright runs.

\section{Prompts and Artifact Schemas}
\label{app:prompts}

\subsection{Prompt Availability}

Exact rendered prompts from the original interactive Stage~1 sessions were not
recorded. We therefore report the role contract, visible inputs, tools, and
output schema rather than reconstructing those prompts after the fact.
Stage~2 and Stage~3 prompt templates are retained in source, although
campaign-era Stage~2 prompt hashes were not recorded on every early run.
\revise{A prompt inventory recording} available paths, hashes, invocation
modes, and campaign associations \revise{will be released with the code}.

\subsection{Role Contracts}

\paragraph{Initial Test-Suite Generator.}
Given the full task instruction and observable initial state, identify explicit
and implicit requirements and generate at least one complete positive state,
one clean negative, and adversarial near-misses. Labels must be justified from
the task instruction alone. The output is a declarative mutation specification
that a deterministic category adapter materializes.

\paragraph{Initial Verifier Evolver.}
Given the materialized challenge states and their instruction-derived labels,
write a sandbox-compatible \texttt{verify(context)} function. When the
internal gate fails, revise the evaluator from deterministic case-level
feedback while preserving already-correct checks.

\paragraph{Workflow Evolver.}
Given the current alternative pool, per-alternative score statistics,
round history, failed checks, and artifact-inspection diagnosis, propose a
small set of new node alternatives. The output is a JSON object containing a
hypothesis and at most one new alternative per selected node.

\paragraph{Refinement Test-Suite Generator.}
Given the task instruction, current evaluator, strong--weak difference
clusters, and immutable state anchors, generate paired cases that reproduce
and remove one selected difference while holding unrelated state fixed.

\paragraph{Refinement Verifier Evolver.}
Given the current evaluator, cumulative cases, paired factual contrasts, and
deterministic feedback, produce a localized patch targeting the uncovered
constraint. A full rewrite is permitted only when the required localized edit
cannot be expressed safely.

\paragraph{Workflow Executors.}
Given one selected workflow path and access to the category-specific tool
interface, execute the task in an isolated environment. The executor does not
receive evaluator source, external-verifier information, or outputs from other
candidate paths.

\subsection{Challenge-State Schema}

A challenge state contains:

{\small
\begin{verbatim}
{
  "case_id": "...",
  "cluster": "...",
  "expected_pass": true | false,
  "adversarial_description": "...", // optional
  "anchor": "f_plus" | "weak",  // refinement only
  "ops": [
    {
      "type": "...",
      "path": "...",
      "content": "...",             // optional
      "from": "..."                 // optional
    }
  ]
}
\end{verbatim}
}

Initial Stage~1 operations are applied to an isolated copy of the initial
environment. Refinement cases may instead be anchored on \(F^{+}\) or the
current weak state.

\subsection{Evaluator Schema}

Evaluators expose one sandboxed function:
{\small
\begin{verbatim}
verify(context) -> {
  "passed": bool,
  "checks": [
    {
      "check_id": str,
      "description": str,
      "requirement_ids": [str, ...],
      "passed": bool,
      "evidence": {...},
      "severity": "required" | "optional"
    }
  ],
  "summary": str
}
\end{verbatim}
}

The Filesystem evaluator context provides nine read-only operations for file
existence, directory existence, globbing, directory listing, text, JSON, and
CSV reading, text hashing, and file-size inspection. Imports, direct file
opening, code evaluation, subprocesses, network access, and filesystem writes
are prohibited.

\subsection{Workflow-Alternative Schema}

Each active workflow node stores alternatives of the form:
{\small
\begin{verbatim}
{
  "alt_id": "...",
  "prompt_template": "...",
  "system_prompt": "...",             // optional
  "max_react_iterations": 20  // ReAct node only
}
\end{verbatim}
}

A workflow stores the ordered node list, alternatives, protected alternatives,
caps, accumulated per-alternative statistics, and path history. A candidate
path selects one alternative from each active node.

\subsection{State-Difference and Counterfactual Schemas}

A canonical difference cluster records:

{\small
\begin{verbatim}
{
  "kind": "content" | "missing_file" | "extra_file"
          | "missing_dir" | "extra_dir" | "empty_dir"
          | "symlink" | "mode",
  "characteristic": "...",
  "paths": ["...", ...],
  "signature": "..."
}
\end{verbatim}
}

A paired counterfactual uses the same cluster identifier for:

\begin{itemize}
    \item a negative case that reproduces the selected difference; and
    \item a positive control that removes only that difference.
\end{itemize}

The label-consistency gate validates their expected verdicts, while the
coverage gate confirms that the pair actually toggles the targeted cluster.

\subsection{Cumulative Regression and Loop State}

The cumulative regression artifact records:

{\small
\begin{verbatim}
{
  "cumulative_freeze_decision":
      "FREEZE" | "REJECT",
  "all_prior_weak_states_rejected": true | false,
  "cumulative_weak_states": [...],
  "added_check_ids": [...],
  "removed_check_ids": [...],
  "case_results": [...]
}
\end{verbatim}
}

The Stage~3 loop state stores evaluator hashes, weak-state tree hashes,
per-round differences and telemetry, the set of seen and covered clusters,
stagnation decisions, and the final terminal reason. The run manifest binds
this lineage to a run identifier, source commit, and frozen strategy.

\end{document}